\pdfoutput=1

\documentclass[%
 aps,
rsi,%
 amsmath,amssymb,
 reprint,%
 showpacs,preprintnumbers,floatfix
]{revtex4-1}

\usepackage{graphicx}
\usepackage{dcolumn}
\usepackage{bm}
\usepackage{hyperref}
\usepackage{caption}
\usepackage{subcaption}
\usepackage{graphics}
\usepackage{float}
\usepackage{epsfig}
\usepackage{setspace}
\usepackage{hyperref}
\usepackage{flafter}
\usepackage{placeins}
\usepackage[utf8]{inputenc}

\begin{document}

\preprint{PITT-PACC-1516}

\title{Transversality of gluon mass generation through an effective loop expansion in covariant and background field gauges}

\author{F.A.Machado}
\email{faa68@pitt.edu}
\affiliation{
Department of Physics and Astronomy, University of Pittsburgh, Pittsburgh, PA 15260, USA.
}%


\begin{abstract}

Gluon mass generation is investigated for 4-dimensional $SU(N)$ Yang-Mills in conventional covariant and in background field gauges within an effective description that,
through a parameterization, can be regarded as a massive gluon model, or as a Nambu-Jona-Lasinio-like expansion around a
massive leading order while preserving the Yang-Mills Lagrangian.
We employ a renormalization scheme that introduces the ratio of the gluon mass parameter $m$ to the saturation value of the gluon propagator. This, along with the mass $m(\mu)$ and the strong coupling $\alpha_s(\mu)$, provided the fit parameters for comparison with $SU(3)$ lattice results
renormalized at the scale $\mu$.
We obtain two types of solutions with satisfactory fits.
Within the proposed expansion, we show the conditions under which it is possible to obtain an exactly vanishing longitudinal self-energy for any gauge parameter $\xi$ in the background field case.
However, such a result in conventional covariant gauges is unattainable by the given expansion as it is, indicating that more sophisticated versions with dressed vertices are likely necessary.
We argue that the conditions under which transversality takes place are consistent with Schwinger-Dyson results for the ghost dressing function also.

\end{abstract}

\maketitle
%
%

\section{Introduction} \label{sec:intro}

It is a well-known fact that Quantum Chromodynamics (QCD) can be described perturbatively in the ultraviolet and that,
as the energy scale decreases, one must resort to either nonperturbative or effective descriptions.
To make sense of this strongly interacting, multi-particle quantum dynamics, and how it leads to a colourless spectrum, 
is in fact a highly non-trivial task, and one of the main open problems in Physics.

Even at zero temperature, some sort of phase transition must occur, as the degrees of freedom of QCD are not the observed ones,
and an actual description of how the theory undergoes confinement is a major, yet long-term goal in the area.
Progress has been steadily made on the investigation of the infrared (IR) behavior of the QCD degrees of freedom, particularly their correlation functions,
through a variety of approaches, e.g. lattice calculations \cite{Cucchieri:2006tf,Ilgenfritz:2006he,Oliveira:2006yw,Cucchieri:2008qm,Mendes:2008ux}, Schwinger-Dyson equations (SDEs) \cite{Aguilar:2008xm,Aguilar:2012rz,Aguilar:2013xqa,Binosi:2014aea,Meyers:2014iwa,Huber:2012zj,Blum:2014gna,Alkofer:2000wg}, nonperturbative quantization
\cite{Dudal:2011gd,Sorella:2011tu}, nonperturbative renormalization group \cite{Weber:2011zzd,Fischer:2009tn}, and effective Lagrangians \cite{Pelaez:2014mxa,Gracey:2014dna}.

While chiral symmetry breaking and the significant dynamical masses of quarks have long been known to be a key feature of nonperturbative QCD,
during the past decade it has been established that dynamical mass generation also happens in (pure) Yang-Mills (YM).
Suggestions that gluons might develop a massive behaviour date further back \cite{Smit:1974je,Cornwall:1981zr}, while early SDE calculations \cite{Mandelstam:1979xd,Brown:1988bm}
did not consider those. As technical power increased, SDE studies pointed to two types of solutions for the YM propagators: the scaling, and the decoupling ones \cite{Fischer:2008uz}.
The former consists in an IR enhanced ghost propagator and an asymptotically zero gluon propagator as the momentum $p^2 \to 0$, while in the latter scenario the ghost dressing
(i.e. its propagator $\tilde{G}(p^2)$ times $p^2$) is finite and the gluon propagator remains finite and non-zero.
As other approaches, especially lattice \cite{Bogolubsky:2009dc} and the Refined Gribov-Zwanziger \cite{Dudal:2008sp}, presented results in agreement with the 
decoupling solution, it has become the most widely accepted one.

Of course, the collection of results on these Green functions is gauge-fixed.
There are investigations, for instance, in Coulomb \cite{Szczepaniak:2001rg} and in maximal Abelian gauges \cite{Mendes:2006kc}, and so far the most explored one
is the Landau gauge \cite{Aguilar:2008xm,Fischer:2003rp,Blum:2014gna,Pelaez:2014mxa,Siringo:2015wtx}. In fact, one can safely say that the decoupling solution is well established for Landau gauge YM in $D=4$ and $D=3$ dimensions \cite{Cucchieri:2007rg,Cucchieri:2008fc}, while the $D=2$ theory seems to agree with the scaling solution \cite{Cucchieri:2009zt}.

The decoupling solution means that the gluon propagator is damped in the IR, and this damping can be described by a gluon mass \cite{Aguilar:2015bud,Binosi:2012sj,Oliveira:2010xc},
which is dynamically generated and intrinsically dependent on the momentum, so the propagator nonperturbatively undergoes the transition
\begin{equation} \label{eq:dyngenmass}
\frac{1}{p^2} \longmapsto \frac{1}{p^2+m_g^2(p^2)} ~.
\end{equation}

Dynamical mass generation itself shows that quantum fluctuations might generate behaviours that are unattainable
 by a perturbative expansion from the Lagrangian, which assumes both (1) a \emph{finite}-order truncation in the usual perturbative expansion, and 
 (2) that it is done around the perturbative, ultraviolet vacuum.
However, confinement seems to somehow involve a transition of the vacuum state of the theory: 
whether the leading feature is non-trivial vacuum condensates \cite{Verschelde:2001ia}, dynamical mass generation \cite{Dudal:2003by,Ayala:2015axa}, Gribov copies \cite{Dudal:2009xh},
 or topological configurations such as instantons or vortices \cite{Dudal:2015khv,Quandt:2010yq}, and whether or not these features are related to
 each other is currently object of investigation, in YM and QCD, in zero and finite temperatures.
 The understanding of the nonperturbative vacuum is likely one essential step towards actual description and understanding of confinement.
 
 There are, also, proposals to circumvent assumption (1) above by constructing improved versions of the usual perturbative QCD series
 \cite{Shirkov:2006gv,Pagels:1979hd,Kneur:2013coa}, with e.g. a scheme for better convergence \cite{Shirkov:2006gv} or the incorporation of
 nonperturbative effects \cite{Pagels:1979hd}, which include certain phenomenological calculations employing dressed gluons with an effective mass 
 \cite{Natale:2006nv,Natale:2009uz}. Other approaches, more directed to the nonperturbative domain,
 introduce nonperturbative quark \cite{Chang:1982tw,Yamada:1992qz,Barducci:1987gn} or gluon \cite{Bieletzki:2012rd,Pelaez:2014mxa,Siringo:2015aka}
 masses into loop expansions.
 
 However low (yet non-zero) the gluon mass scale is, it does spoil unitarity in such models,
 which is currently an open problem if one is to compute scattering processes. As argued in \cite{Gracey:2014dna}, these non-unitary models can be regarded as an effective, leading description within a given range
of applicability, which might exclude the deep ultraviolet. Perhaps, concerning unitarity,
it comes down to the point that the gluon mass is \emph{intrinsically dynamical}, going to zero sufficiently fast at high energies \cite{Binosi:2012sj,Aguilar:2013hoa}, and whose precise description would inevitably be non-local.

Although the Curci-Ferrari (CF) \cite{Curci:1976bt,Tissier:2011ey} model is not unitary \cite{Curci:1976kh}, one of its extensions \cite{Delbourgo:1981cm,Serreau:2015yna} is both unitary and renormalizable (the CF Lagrangian and the local mass term $m^2A^2$ are also renormalizable \cite{Dudal:2003np}),
and has been recently applied to the study of YM correlators in (non-linear) covariant gauges \cite{Serreau:2015yna}.
The latter is one of the recent efforts \cite{Aguilar:2015nqa,Huber:2015ria,Bicudo:2015rma,Capri:2015pja,Siringo:2015gia} on the study of covariant gauges \cite{Cucchieri:2011pp,Giusti:1996kf}.

In fact, most of the evidence for a gluon mass comes from Landau gauge calculations, and it is presently an open question 
how dynamical mass generation works for gluons in other gauges, particularly the $R_\xi$ class.
So far, lattice \cite{Bicudo:2015rma} and Refined Gribov-Zwanziger \cite{Capri:2015ixa} approaches obtain a tree-level form for the 
longitudinal component of the gluon propagator at the nonperturbative level, thus respecting the corresponding Slavnov-Taylor identity (STI).
Among the SDE calculations, until now only the ones within the framework of the Pinch Technique (PT) satisfy transversality of the gluon self-energy within each truncation \cite{Binosi:2003rr,Binosi:2009qm}.
For SDE calculations in the usual framework, as well as for effective models with massive gluons \cite{Pelaez:2014mxa,Siringo:2015aka},
there is no guarantee that the longitudinal component will not receive corrections, which means the truncation (or the model) violates this essential symmetry.
In fact, as we will show, even in Landau gauge a non-zero longitudinal self-energy is generated in massive gluon expansions.

The problem with transversality is twofold. From first principles, the longitudinal component $1/\xi$ in linear covariant gauges originates from the width of the gauge-fixing distribution \cite{Huber:2010ne}, and it remains unaffected due to BRST \cite{Becchi:1975nq} symmetry -- both in its conventional form and in its extension to the nonperturbatively quantized Gribov-Zwanziger action \cite{Capri:2015ixa}.
Second, from the practical side, a nonzero longitudinal self-energy might generate extra longitudinal terms or spurious polarization states, making further applications of such expansions potentially inaccurate or ambiguous.

In other words, generating a nontrivial longitudinal dynamics breaks not only BRST symmetry, but also the fact that gluons are fundamentally two dynamical degrees of freedom. In this sense, this problem is present even in Landau gauge, where the tree-level form $1/\xi$ projects the longitudinal propagator to zero unless its self-energy has a singularity in $\xi=0$ (see Eq.(\ref{eq:ldressing})), while the self-energy, and therefore the proper two-point function (i.e. the inverse propagator) are not identically zero. Ultimately, the problem of transversality is connected to describing the dynamically massive gluons as still having two and only two independent polarization states, in any gauge.

The present work aims to investigate transversality of the gluon self-energy in a loop expansion with massive gluons, similar to the ones employed in \cite{Tissier:2011ey,Siringo:2015aka}.
Such an expansion can be regarded, on one hand,
as a prototype for an improved loop expansion to be implementable in usual perturbative QCD calculations, 
which could extend their validity deeper into the IR and with faster convergence \cite{Shirkov:2001sm,Dokshitzer:1998nz}.  In this aspect, technical simplicity is desired.

On the other hand, it can be useful as a simple tool to access the IR domain of the theory, and probe quantities and processes
that belong or relate to this domain. Thus it could provide a link from the ultraviolet (UV) to the IR domain, while being a tool to explore the IR itself.

The question the present work aims to answer is: is there a form for an input gluon propagator, and a given sort of expansion that
can yield a purely transverse gluon self-energy, by dressing only the gluon propagators (meaning, no dressed vertices)? How well do the results agree with other approaches?

As an exploratory calculation, we then propose a dressed-gluon, Nambu-Jona-Lasinio(NJL)-like \cite{Nambu:1961tp,Chang:1982tw} expansion in $R_\xi$ gauges. We apply it to the gluon and ghost propagators in 4-dimensional pure YM, which can be compared to reported results from SDEs \cite{Aguilar:2015nqa,Huber:2015ria} and lattice \cite{Bicudo:2015rma}. For the purposes of this paper, the matters of analytic continuation and unitarity will not be touched, all analysis being concerned with Euclidean space solutions for the given correlators.

After detailing the approach in Section \ref{sec:proposal}, in Section \ref{sec:gluon} we describe the implementation of self-consistency as a renormalization condition for the mass counterterm. In \ref{sec:cg} we present the results for conventional covariant gauges (CG), which are compared to lattice results from \cite{Bicudo:2015rma} and explored from there. In \ref{sec:bfg}, we analyze results for background field gauges (BFG), which will lead to Subsection \ref{sec:transv} on the matter of transversality of the gluon self-energy. 
Then, we present the corresponding results for the ghost propagator in \ref{sec:ghost}, in \ref{sec:rg} we make some preliminary observations on the renormalization group (RG) treatment and possible features of the paper's proposal, and from that we proceed to concluding remarks in \ref{sec:concluding}. 

\section{The NJL-like proposal} \label{sec:proposal}

It is known that effective models with massive gluons can successfully fit lattice data for YM Green functions \cite{Pelaez:2013cpa}.
One can take these sorts of approaches as effective models of YM in the IR \cite{Tissier:2011ey},  or else as effective expansions \cite{Siringo:2015wtx}, in some kind of resummation or reorganization of the series.
The present investigation contains both of these types of approach as particular cases, through the parameter $\lambda$ given below.

Starting with the $R_\xi$ gauge-fixed YM Lagrangian,
\begin{eqnarray}
{\mathcal L}_{\text{YM}} &=&-\frac{1}{4}F^2 -\frac{1}{2\xi }\left( \partial . A\right)^2  + \left( \partial \bar{c} \right). \left(D c\right)  \notag
\\
&=& {\mathcal L}_0 + {\mathcal L}_{{\text i}} +{\mathcal L}_{\text{ct}} ~, \label{eq:basiclag}
\end{eqnarray}
where $F$ is the gluon field strength, $A$ the gluon field, $\xi$ the gauge parameter, $D$ the covariant derivative, $c(\bar{c})$ the (anti-)ghost field,
and the low dots and the squares represent Lorentz and color tracing. On the second line (\ref{eq:basiclag}) we write it as the sum of kinetic terms,
interaction ones, and counter-terms, respectively.

Then, inspired by the expansions of e.g. \cite{Nambu:1961tp,Chang:1982tw} around a nonperturbative vacuum, we add a gluon self-energy term, ${\mathcal L}_{\text{se}}$, and an interaction term $-\lambda{\mathcal L}_{\text{se}}$,
so the new Lagrangian is:
\begin{eqnarray}
{\mathcal L} &=& {\mathcal L}_{\text{YM}} + (1-\lambda) {\mathcal L}_{\text{se}} \notag \\
&=& \left( {\mathcal L}_0 + {\mathcal L}_{\text{se}} \right) + \left( {\mathcal L}_{{\text i}} 
-\lambda {\mathcal L}_{\text{se}} \right) +{\mathcal L}_{\text{ct}} ~.  \label{eq:lambdalagrangian}
\end{eqnarray}
The second line expresses that we shall treat ${\mathcal L}_{\text{se}}$ now as part of the free Lagrangian,
while the term $-\lambda{\mathcal L}_{\text{se}}$ is to be treated as a higher order interaction one.
So, to order $\alpha_s$, the corresponding counterterm added will be
\begin{equation} \label{eq:nolambdact}
(1-\lambda)\delta \mathcal{L}_{se} = \delta \mathcal{L}_{se} + \mathcal{O}({\alpha_s^{2}}) ~.
\end{equation}
The idea is that the nonperturbative dynamics, and all its possible features mentioned in Section \ref{sec:intro}, is \emph{effectively} taken into account by the gluon self-energy.

We would like $\mathcal{L}_{\text{se}}$ to mimic the nonperturbative gluon self-energy.  According to many results in Landau gauge \cite{Aguilar:2008xm,Aguilar:2015bud,Bicudo:2015rma}, the dynamical gluon mass is purely transverse. However, as it will be explicitly shown, a massive model in Landau gauge does generate a longitudinal self-energy altogether -- although the propagator is zero, the self-energy is not. 

So, in order to explore the mass generation in $R_\xi$, we consider not only a transverse, but also a longitudinal mass, through the following self-energy Lagrangian:
\begin{equation} \label{eq:lse}
{\mathcal L}_{\text{se}} = - \frac{1}{2} A^\mu \left( m_T^2 \left( \delta_{\mu\nu} - \partial_\mu \partial_\nu  \partial^{-2} \right) 
+  \frac{m_L^2}{\xi} \partial_\mu \partial_\nu  \partial^{-2} \right) A^\nu ~,
\end{equation}
where we assume that the non-local operator $ \partial_\mu \partial_\nu \partial^{-2}$ is well-defined for the gluon states in question. Were we aiming at a more complete, properly formal approach, we could introduce auxiliary fields and, with a new local Lagrangian, explore the symmetries and possibly write extended versions of Slavnov-Taylor identities, as some approaches do \cite{Capri:2015ixa,Wschebor:2007vh}. 
 However, since the objective so far is to explore the properties and effects of this dressed expansion to 1-loop, we will restrict ourselves to employing the corresponding gluon propagator, given by
\begin{equation}
G_{(0)}(p) = \frac{1}{p^2+m^2} \perp_{\mu\nu}(p) + \frac{\xi}{p^2+m^2 r} \parallel_{\mu\nu}(p) ~,
\end{equation}
where $\perp_{\mu\nu}(p)=\delta_{\mu\nu}-\parallel_{\mu\nu}(p)$, $\parallel_{\mu\nu}(p)=p_\mu p_\nu /p^2$, and we write $r:=m_L^2/m_T^2$ and $m_T=:m$.

Two values of $r$ will be of our concern: $r=0$ and $r=\xi$. The former corresponds to a purely transverse ${\mathcal L}_{\text{se}}$, and therefore a massive transverse gluon propagator together with the usual tree-level longitudinal part. 
The latter corresponds to the local $m^2 A^2$ Lagrangian term, which leads to the usual massive vector boson propagator that describes three independent polarization states.

Moreover, based on (\ref{eq:nolambdact}) and (\ref{eq:lse}), the counterterms corresponding to the $A_\mu A_\nu$ term are:
\begin{eqnarray}
{\mathcal L}_{\text{ct;}AA}&=& -\frac{1}{2}\delta_Z A^\mu \left( \partial_\mu \partial_\nu  - \delta_{\mu\nu} \partial^2 \right) A^\nu   \notag \\
&&+\frac{1}{2} m^2 \delta_T A^\mu \left( \delta_{\mu\nu} - \partial_\mu \partial_\nu  \partial^{-2} \right) A^\nu \notag \\
&&+\frac{1}{2} m^2 \delta_L A^\mu \left( \partial_\mu \partial_\nu  \partial^{-2} \right) A^\nu ~, \notag
\end{eqnarray}
where $\delta_T$ and $\delta_L$ could in principle depend on $r$. As we will show, it turns out they do not.

So, the picture is the following: for $\lambda=0$, we have a massive gluon model, while by keeping $-\lambda{\mathcal L}_{\text{se}}$ 
we can treat it as of ${\mathcal O}(\alpha_s)$, and then properly set $\lambda=1$ in the 1-loop result, 
in a similar manner to NJL's approach and some of its extensions \cite{Chang:1982tw,Yamada:1992xg}.
Keeping $\lambda$ as a parameter allows us to explore how well can these two approaches work.

\section{Gluon propagator} \label{sec:gluon}

We write the gluon propagator, decomposed into transverse and longitudinal components, as 
\begin{equation}
G_T(p^2) \perp_{\mu\nu} + \, G_L(p^2) \parallel_{\mu\nu} ~.
\end{equation}
Denoting $\bar{\alpha}:=N\alpha_s/48\pi$ and $s:=p^2/m^2$, we can write our 1-loop result as
\begin{eqnarray}
G_T^{-1} &=& p^2 \left( 1 +\bar{\alpha} f_T(s,r)+ \delta_Z \right) + m^2 (1-\lambda+ \delta_T) ~, \label{eq:gt} \\ 
G_L^{-1} &=& p^2 \left( \frac{1}{\xi} +3\bar{\alpha} f_L(s,r) \right) + m^2 (1-\lambda) \frac{r}{\xi} +m^2 \delta_L ~, \label{eq:gl} 
\end{eqnarray}
where $f_T$ and $f_L$ are shown in Appendix \ref{sec:appendix} for both CG (\ref{sec:appcg}) and BFG (\ref{sec:appbfg}).
Before we consider their specifics, we should deal with the presence of the $m^2$ counterterms, making it necessary to impose two renormalization conditions.
In principle there could be three, one for $\delta_L$ as well. However, as one can see from all results in Appendix \ref{sec:appendix}, the $m^2/(D-4)$ pole in each $G_T$ is the same as in the corresponding $G_L$. In other words, the $m^2$ UV divergences are diagonal, and therefore we can renormalize $G_L$ by setting $\delta_L = \delta_T$, while we present below some reasons for choosing a given $\delta_T$ scheme.

\subsection{Self-consistency as a renormalization scheme}

First, we verified for our loop results that, in general, different renormalization schemes ($\overline{\text{MS}}$ included) yield, up to an overall multiplicative factor, consistent results provided the initial values of parameters and renormalization scale are properly chosen.

We definitely do not have the same degree of arbitrariness for renormalization as we have in perturbative QCD, and we do not have pole mass conditions, since gluons are confined. One reference we do have for the gluon mass is the lattice computations. Specifically, the lattice saturation value $G_T(0)$ can be directly related to the saturation of the dynamical mass, $m_g(0)$, through Eq.(\ref{eq:dyngenmass}). Thus, one can take $m_g(0)$ from the lattice as a sort of boundary condition, which will renormalize the $m^2$ UV divergence.

At this point, we recall that the self-consistency requirement in NJL's work \cite{Nambu:1961tp} implied a relation between the mass and the cutoff, and note that by taking a renormalization condition for $\delta_m$ from self-consistency, the requirement will thus determine the dependence of the effective gluon mass $m$ with the renormalization scale $\mu$. That is, 
$$
\gamma_m = \frac{\mu}{m^2} \frac{\mathrm{d}m^2}{\mathrm{d}\mu} 
=  - \frac{\mu}{m^2} \frac{\mathrm{d}\delta_m}{\mathrm{d}\mu} + \mathcal{O}(\alpha_s^2) \Longrightarrow m^2(\mu)
$$
is directly given by a self-consistency condition.

Now, concerning what would be a reasonable self-consistency condition within the present proposal, we recall that similar NJL-like approaches were employed to study dynamical quark mass generation \cite{Chang:1982tw,Yamada:1992xg}. Ref.\cite{Chang:1982tw}, for instance, imposes that the quark self-energy is zero in the limit $p^2/M^2 \to 0$, where $M$ is their quark version of our gluon mass $m$.

As explained above, our choice will concern the saturation of the gluon propagator. Specifically, for the transverse gluon self-energy:
\begin{equation}
\Pi_T (p,m) = p^2 \left[ \bar{\alpha} f_T(s,r) +\delta_Z \right]+ m^2 \delta_T  - \lambda m^2  ~,
\end{equation}
we can impose, as self-consistency condition:
\begin{equation} \label{eq:freezingcondition}
\Pi_T(0,m) = z_0 m^2 ~,
\end{equation}

Condition (\ref{eq:freezingcondition}) states that the ratio $m_g(p^2=0)/m$ between the saturation value of the dynamical gluon mass and the transverse mass parameter of the model will equal $\sqrt{1+z_0}$.

In the point of view where the present expansion is an approximation of the SDEs with bare vertices and $m_g(p^2)$ is approximated as a constant $m$, it is fair to take $m$ as some sort of average of $m_g(p^2)$. Since the latter is monotonically decreasing \cite{Aguilar:2011ux}, one expects that $m_g(0) \geq m > 0$, i.e. $z_0 \geq 0$.
However, we will allow $z_0$ to be negative, in which case one can interpret $m$ as an effective parameter that would account not only for the dynamical mass, but possibly other IR features, whether they would be condensates, vortices, or something else.

From (\ref{eq:gt}) and (\ref{eq:freezingcondition}), and writing $\delta_T = \delta_m + \delta_Z $, we obtain:
\begin{equation} \label{eq:selfconsistency}
m^2 \left( -z_0 + \delta_m + \delta_Z -\lambda + \bar{\alpha}  \lim_{s \to 0}s \, f_T(s,r) \right) =0 ~,
\end{equation}
and, like in \cite{Nambu:1961tp}, we note $m=0$ as a possible solution, and as long as
$$
\lim_{p^2 \to 0}p^2 f_T(s,r) 
$$
is finite, there can also be non-trivial, $m\neq 0$ solutions.

So, employing (\ref{eq:selfconsistency}) as a renormalization condition, we have:
\begin{equation} \label{eq:deltamscheme}
\delta_m (\mu) = z_0 + \lambda - \delta_Z(\mu) - \bar{\alpha} \lim_{s \to 0} s \, f_T(s,r) ~.
\end{equation}

Since $\lambda$ is absorbed by $\delta_m$, $G_T$ is insensitive to $\lambda$, thus only $G_L$ will be able to tell us if there is one preferable value for it, either $0$ or $1$. We should remark, though, that this cancellation occurs in many other schemes, such as fixing $G_T(\mu)$ and $G'_T(\mu)$, or fixing $G_T$ at two scales, for example.

Also, we note that a gauge-independent $z_0$ means we would be imposing the saturation value to be the same for all values of $\xi$. However, results from both lattice \cite{Bicudo:2015rma} and SDEs \cite{Aguilar:2015nqa,Huber:2015ria} indicate that $G_T(0)$ would depend on $\xi$ \cite{Quadri:2014jha}. So we adapt our self-consistency condition to match the gauge-dependence evidenced by those, by making $z_0 \mapsto z_\xi(\xi)$ in the following way. Ref. \cite{Aguilar:2015nqa} obtains solutions with
$$
\frac{m_g^2(p^2=0,\xi)}{m_g^2(p^2=0,\xi=0)} = a(\xi) = 1+a_1 \xi ~,
$$
which, since $1+z_\xi(\xi) = m_g^2(p^2=0,\xi)/m^2$, implies
$$
\frac{1+z_\xi(\xi)}{1+z_\xi(0)} =  1+a_1 \xi ~,
$$
which in turn implies 
\begin{equation} \label{eq:zxi}
z_\xi(\xi) = z_\xi(0)+ a_1 \left( 1 + z_\xi(0) \right) \xi ~.
\end{equation}
Then, we can take from \cite{Bicudo:2015rma} the numerical value $a_1 = 0.26$, and implementing (\ref{eq:zxi}) we return to having only one variable $z_\xi(0)=:z_0$, which we call $z_0$ again.

Moving forward, $\delta_Z(\mu)$ is fixed through momentum subtraction scheme, specifically
\begin{equation} \label{eq:deltazscheme}
G_T(\mu) = \frac{1}{\mu^2} ~~~~ \text{    at  }~~ \mu=1 \, \text{GeV} ~.
\end{equation}
The reason for choosing $1$GeV at this point is because around this scale the solutions for distinct renormalization schemes displayed similar behaviors for closer parameter ranges.
Then in a renormalization group analysis, $\mu=1$GeV can be taken for either lower or upper boundary condition, leading respectively to the UV or the IR behavior of the correlators and parameters.

At last, we have a renormalized $G_T$ containing $z_0$, and we shall employ it as a parameter to fit our CG result to available lattice data from \cite{Bicudo:2015rma}, together with $\alpha_s(\mu=1 \, \text{GeV})$ and $m(\mu=1 \, \text{GeV})$.

\subsection{Covariant gauges (CG)} \label{sec:cg}

The covariant gauge results for $f_T$ and $f_L$ in (\ref{eq:gt}) and (\ref{eq:gl}) are given in Appendix \ref{sec:appcg} for arbitrary $r$ and for $r=0$.
One can easily check that $f_L$ is nonzero for $\xi=0$, and even in the sequence of limits $r \to \xi$ and then $\xi \to 0$. 

The poles at $D=4$ are given by:
\begin{eqnarray} \label{eq:cgpoles}
\delta_Z &=& \bar{\alpha} \left( -26 + 6 \xi \right) + ~  \text{finite} ~,\\
\delta_m &=& \bar{\alpha} \left( 35 + 3 \xi \right) + ~ \text{finite} ~.
\end{eqnarray}

\begin{figure*}[h]
\centering
        \begin{subfigure}[h]{0.423\textwidth}
\includegraphics[width=\textwidth]{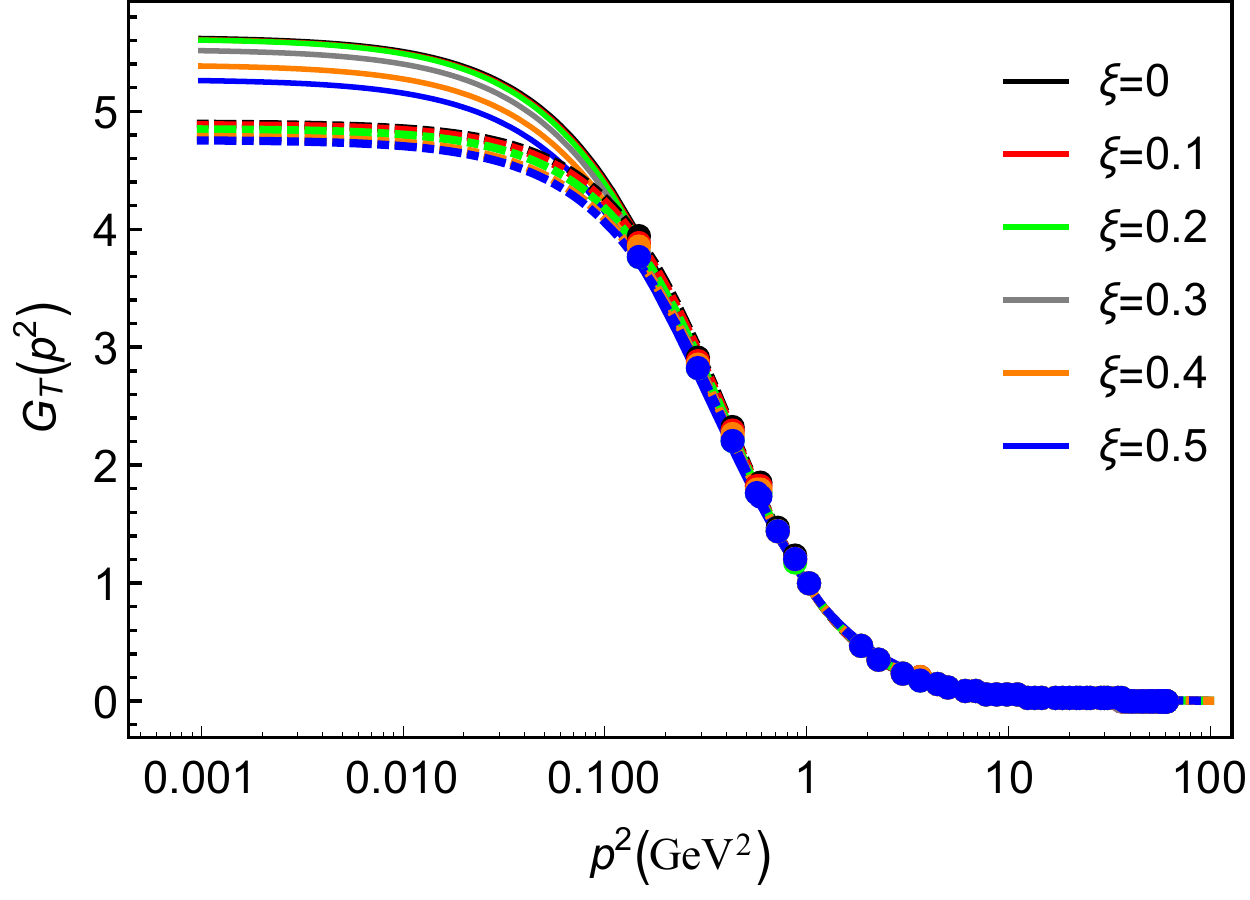}
\caption{Fits to the lattice gluon propagator from \cite{Bicudo:2015rma}.}
 \label{fig:fig1a}
\end{subfigure}
~
        \begin{subfigure}[h]{0.429\textwidth}
\includegraphics[width=\textwidth]{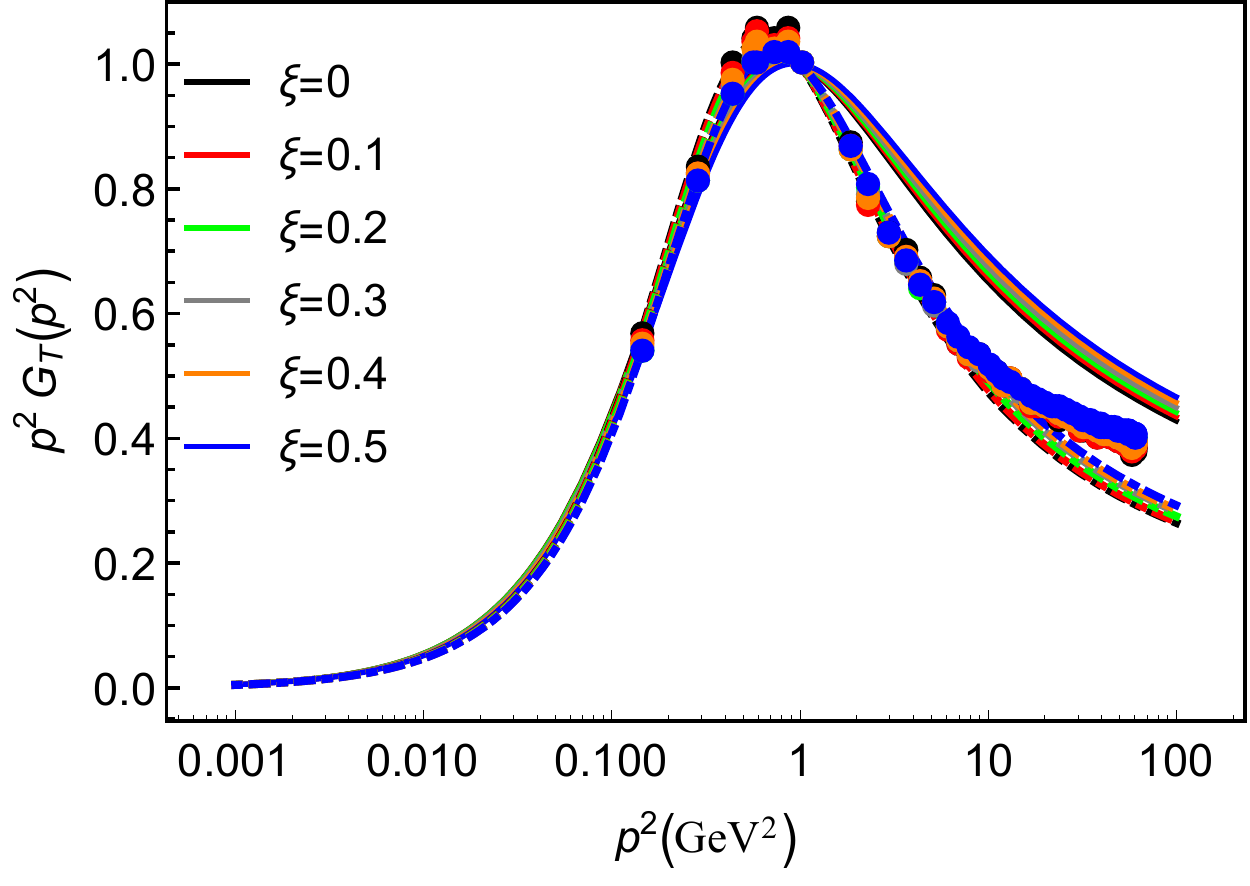}
\caption{Corresponding gluon dressing function.}
 \label{fig:fig1b}
\end{subfigure}
        \caption{Fitted curves for the gluon propagator (\ref{fig:fig1a}), and the dressing function (\ref{fig:fig1b}).  Continuous lines are the $z_0>0$ curves, dashed lines are the fits with $z_0<0$. The points show the results from \cite{Bicudo:2015rma} renormalized to $1$ at $1$ GeV.}\label{fig:fig1}
\end{figure*}

Following (\ref{eq:deltamscheme}) and (\ref{eq:deltazscheme}), we then fitted the lattice data for the covariant gauges, renormalized to $1$ at $1\,$GeV,
obtaining two sets of parameter ranges which correspond to the continuous and the dashed lines in Fig.\ref{fig:fig1}.

The parameters for the continuous lines in Fig. \ref{fig:fig1} belong in the the following ranges:
\begin{eqnarray} \label{eq:z0pos}
z_0 &\in&  [0.20,0.27]; \notag \\
m&=& 374\, \text{MeV;} \notag \\
\alpha_s &=& 0.7; \notag \\
m_g(0) &\in&  [421,435]\, \text{MeV.} \notag \\
\end{eqnarray}
For the dashed lines in Fig. \ref{fig:fig1}, the parameter ranges are:
\begin{eqnarray} \label{eq:z0neg}
z_0 &\in&  [-0.37,-0.27]; \notag \\
m&\in& [529,548]\, \text{MeV;} \notag \\
\alpha_s &\in& [1.48,1.51]; \notag \\
m_g(0) &\in&  [452,457]\, \text{MeV.} \notag \\
\end{eqnarray}

This shows that both types of solutions, namely $z_0>0$ and $z_0<0$, are able to reasonably approximate the lattice results.
Although not displayed in Fig.\ref{fig:fig1a}, the saturation values given by the lattice (results from \cite{Bicudo:2015rma} renormalized to $1$ at $1$ GeV) are closer to the ones for the former case, $z_0>0$, which are the continuous lines in  Fig.\ref{fig:fig1}.
Fig. \ref{fig:fig1b} shows the dressing functions ($p^2 G_T(p^2)$) for these two sets of parameters, also with the corresponding quantity from the
lattice \cite{Bicudo:2015rma}.

\begin{figure*}[h]
\centering
        \begin{subfigure}[h]{0.315\textwidth}
\includegraphics[width=\textwidth]{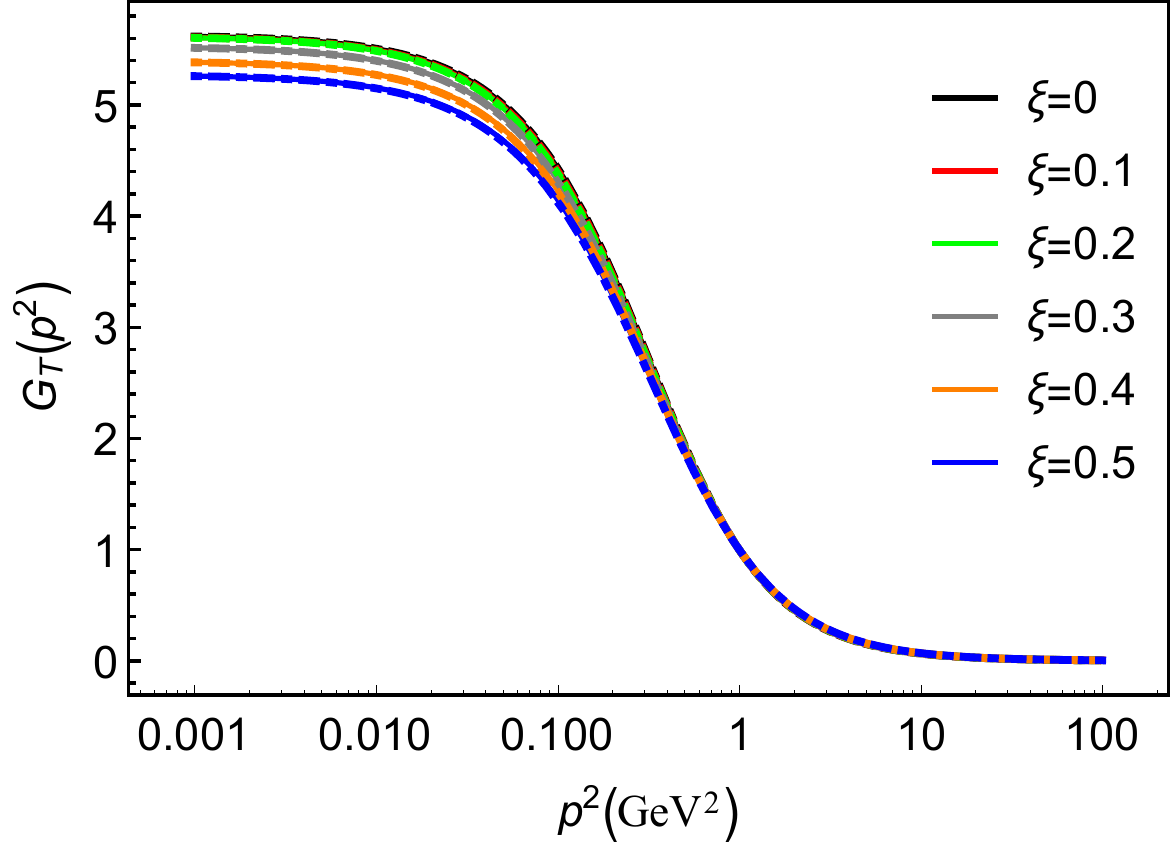}
\caption{}
 \label{fig:2a}
\end{subfigure}
~
        \begin{subfigure}[h]{0.324\textwidth}
\includegraphics[width=\textwidth]{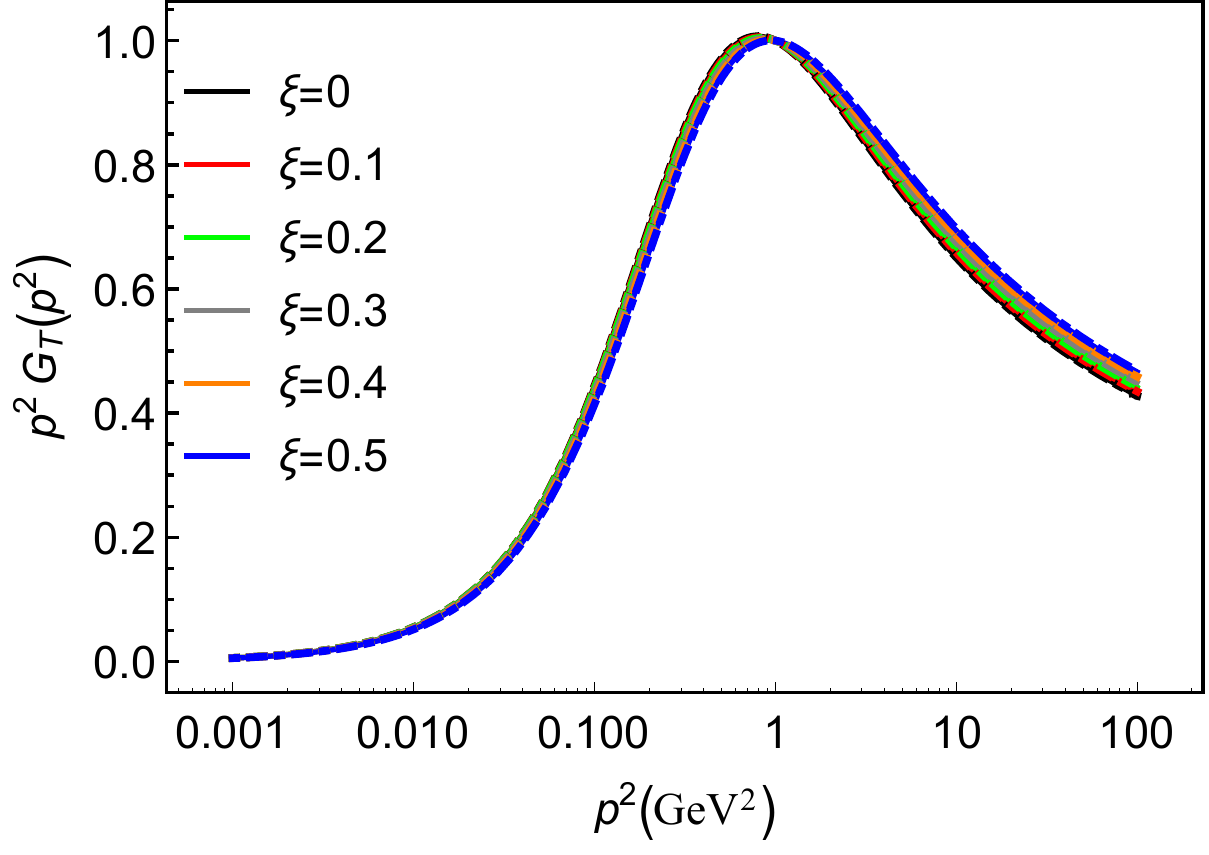}
\caption{}
 \label{fig:2b}
\end{subfigure}
~
        \begin{subfigure}[h]{0.323\textwidth}
\includegraphics[width=\textwidth]{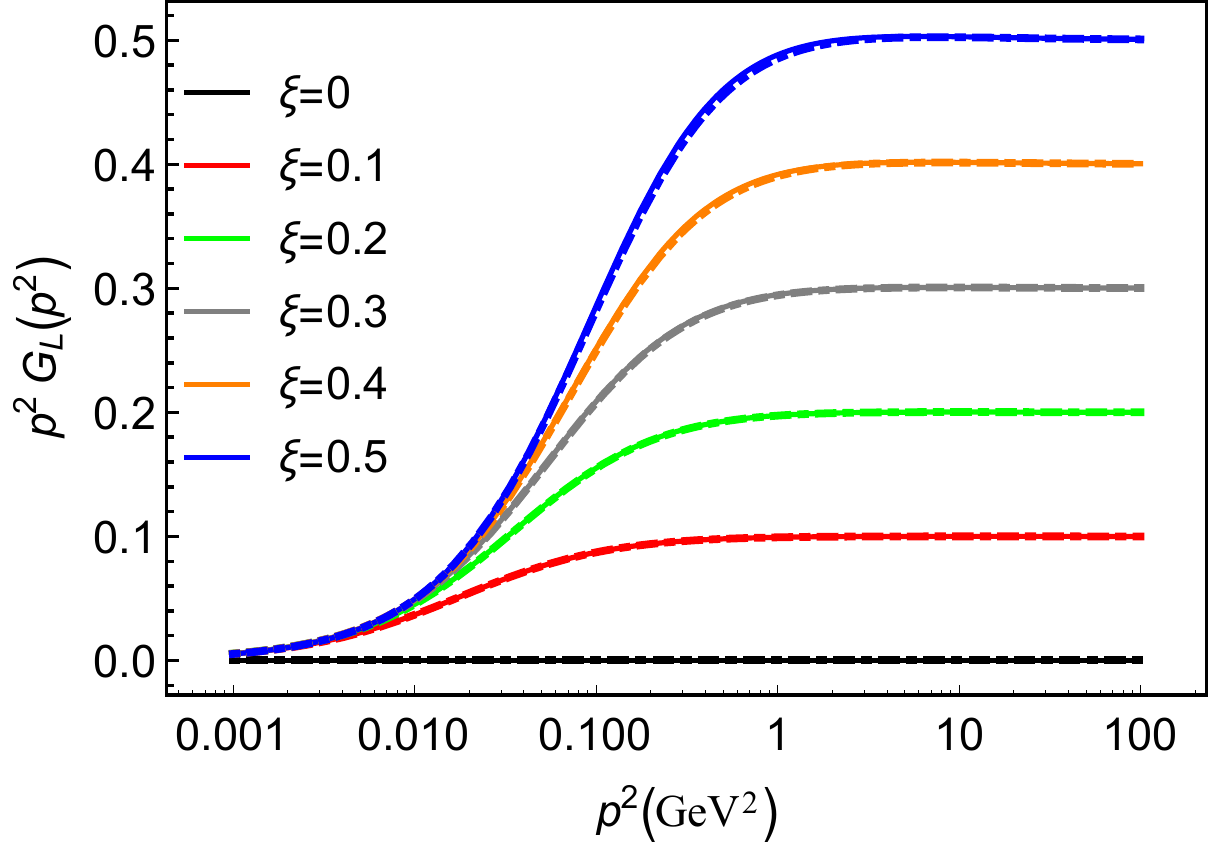}
\caption{}
 \label{fig:2c}
\end{subfigure}
        \caption{CG transverse gluon propagator (a), transverse dressing function (b), and longitudinal dressing function (c), the latter with $\lambda=1$.
Continuous lines correspond to $r=0$, and dashed lines to $r=\xi$.}\label{fig:fig2}
\end{figure*}

In order to explore our results, we will employ certain values for these parameters. Since we are truncating to $\mathcal{O}(\alpha_s)$,
we opt to use the $z_0>0$ set, (\ref{eq:z0pos}), for which a smaller coupling is obtained: $\alpha_s (1\, \text{GeV}) = 0.7$. Fig. \ref{fig:fig2} shows respectively the transverse gluon propagator,
the transverse dressing function, and the longitudinal dressing function (with $\lambda=1$), for CG with $r=0$ (continuous lines) and $r=\xi$ (dashed lines).

One can see that the $r=\xi$ dashed curves are practically identical to their $r=0$ correspondents for each value of $\xi$.
Moreover, Fig. \ref{fig:2c} shows that a nonzero, momentum-dependent longitudinal self-energy is generated, since $p^2 G_L(p^2) \equiv \xi$
is connected to $f_L(s,r) \equiv 0$ through:
\begin{equation} \label{eq:ldressing}
p^2 G_L(p^2) = \xi \left[ 1 + 3 \bar{\alpha} \xi f_L(s,r)+ s^{-1} \left( r(1-\lambda) +\xi \delta_L \right) \right]^{-1} ~.
\end{equation}

\subsection{Background field gauges (BFG)} \label{sec:bfg}

The corresponding plots for the BFG are shown in Fig. \ref{fig:fig3} below.
\begin{figure*}[h!]
\centering
        \begin{subfigure}[h!]{0.315\textwidth}
\includegraphics[width=\textwidth]{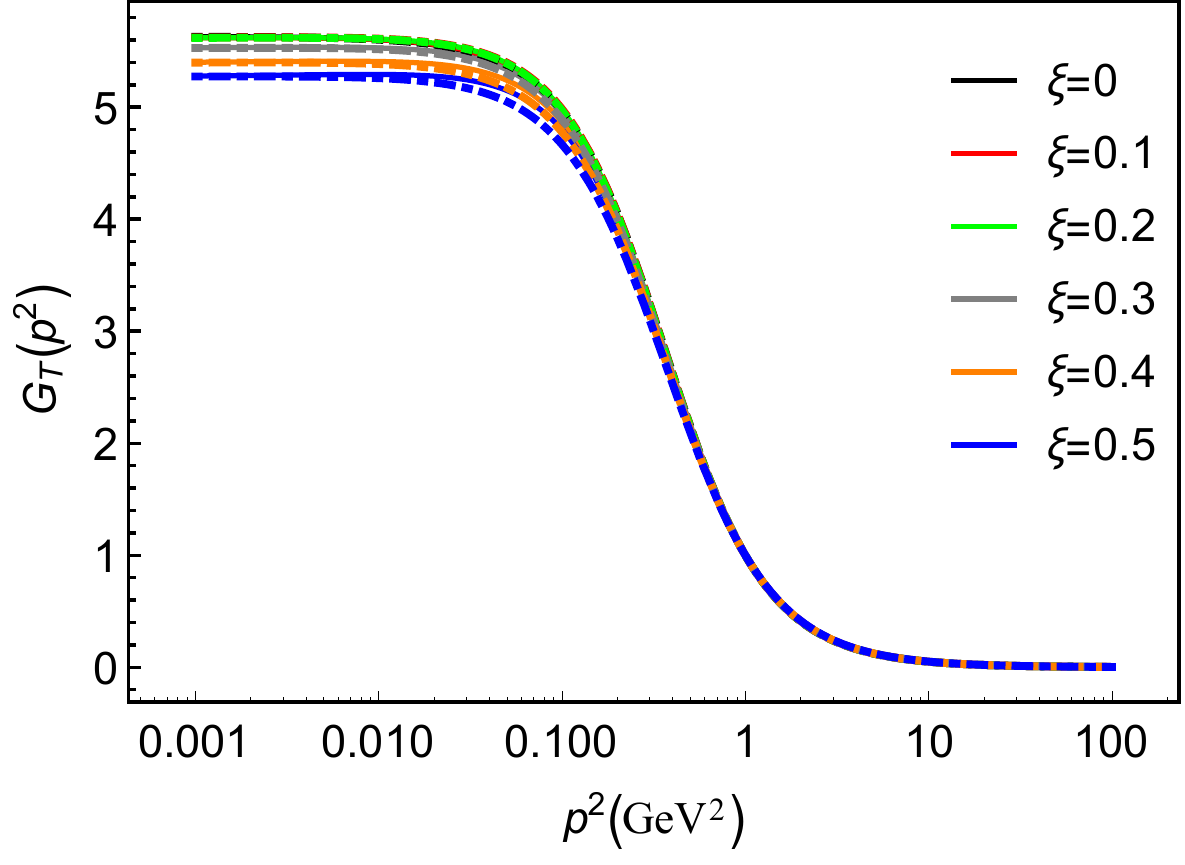}%
\caption{}
 \label{fig:3a}
\end{subfigure}
~
        \begin{subfigure}[h!]{0.324\textwidth}
\includegraphics[width=\textwidth]{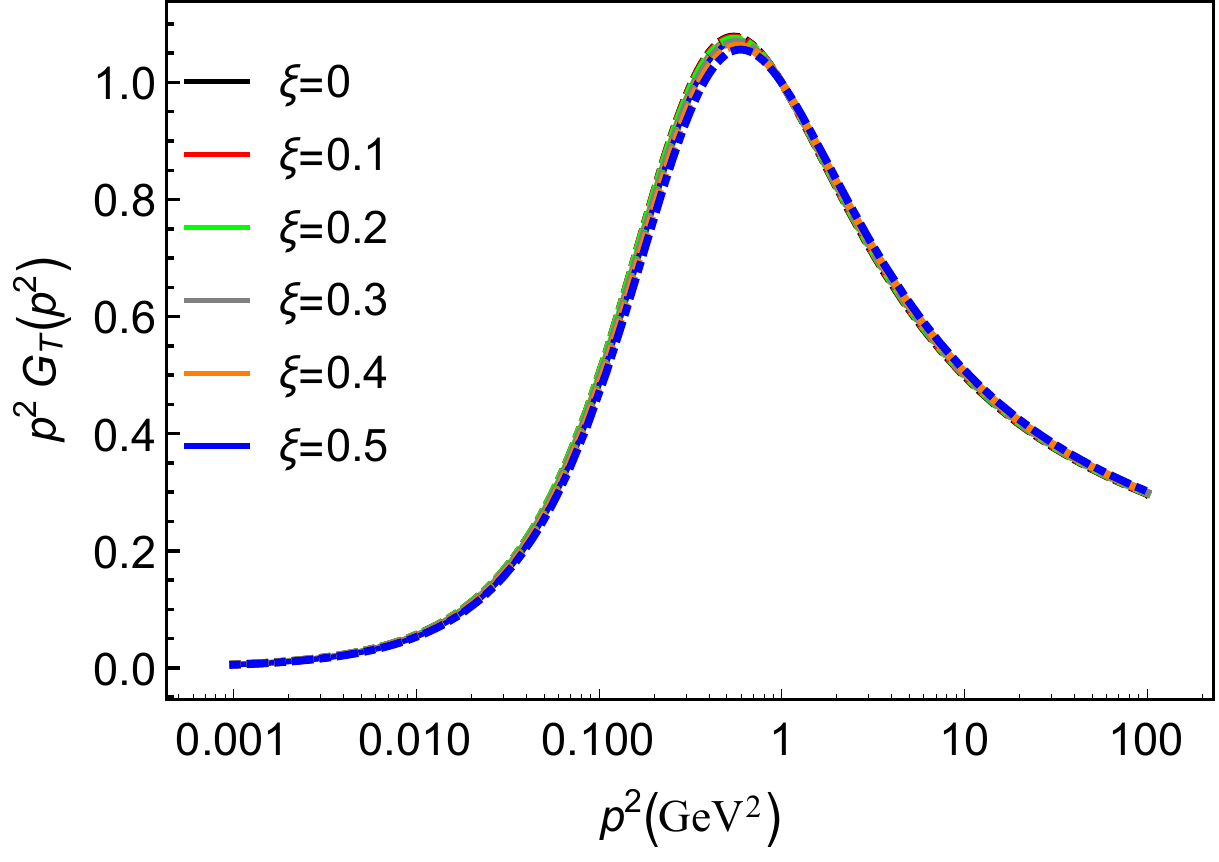}
\caption{}
 \label{fig:3b}
\end{subfigure}
~
        \begin{subfigure}[h!]{0.323\textwidth}
\includegraphics[width=\textwidth]{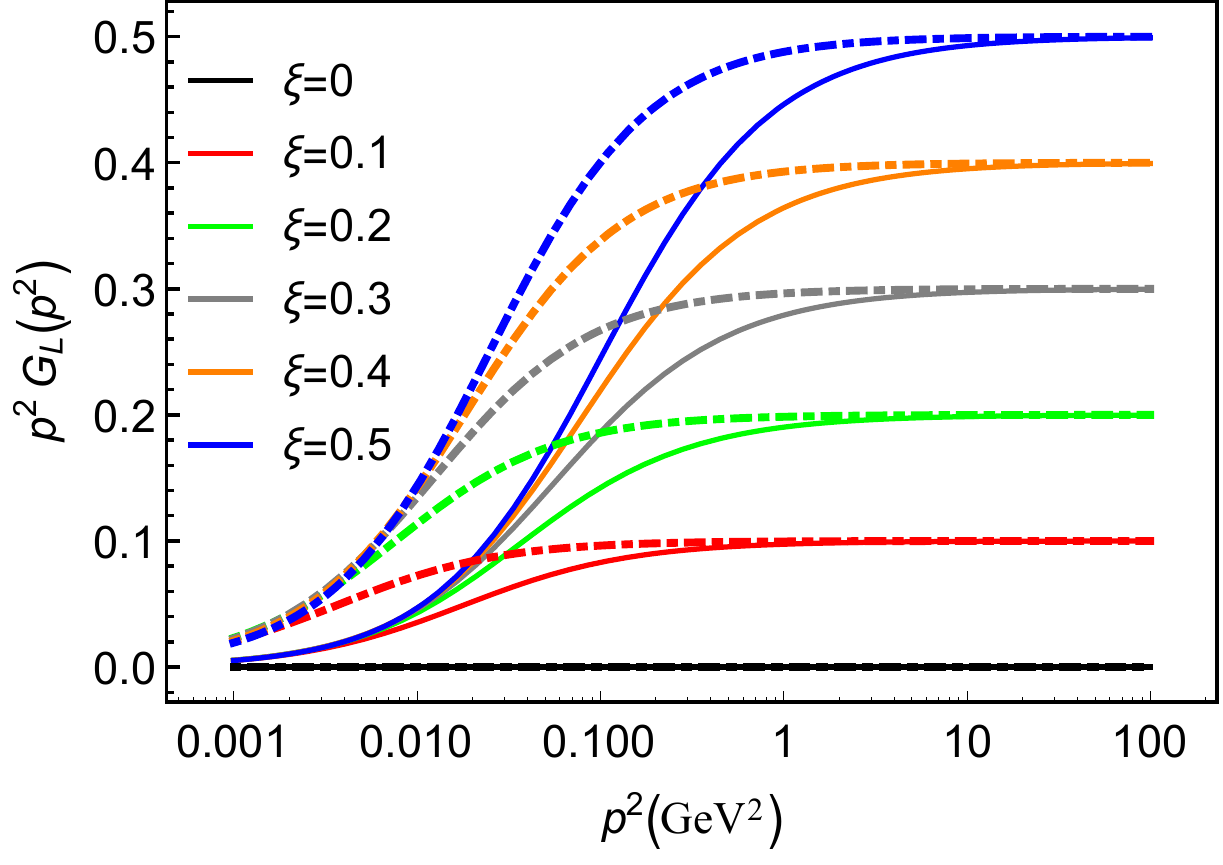}
\caption{}
 \label{fig:3c}
\end{subfigure}
        \caption{BFG transverse gluon propagator (a), transverse dressing function (b), and longitudinal dressing function (c), the latter with $\lambda=1$.
Continuous lines correspond to $r=0$, and dashed lines to $r=\xi$.}\label{fig:fig3}
\end{figure*}
In this case there is some quantitative difference between the $r=0$ and $r=\xi$ cases for the longitudinal dressing, yet they present the same qualitative behavior. 

The poles at $D=4$ are given by:
\begin{eqnarray} \label{eq:bfgpoles}
\delta_Z &=& \bar{\alpha} \left( -44 \right) + ~  \text{finite} ~, \\
\delta_m (r=0) &=& \bar{\alpha} \left( 35 + 9 \xi \right) + ~ \text{finite} ~, \\
\delta_m (r=\xi) &=& \bar{\alpha} \left( 44  \right) + ~ \text{finite} ~.
\end{eqnarray}

\begin{figure*}[h!]
\centering
        \begin{subfigure}[h!]{0.331\textwidth}
\includegraphics[width=\textwidth]{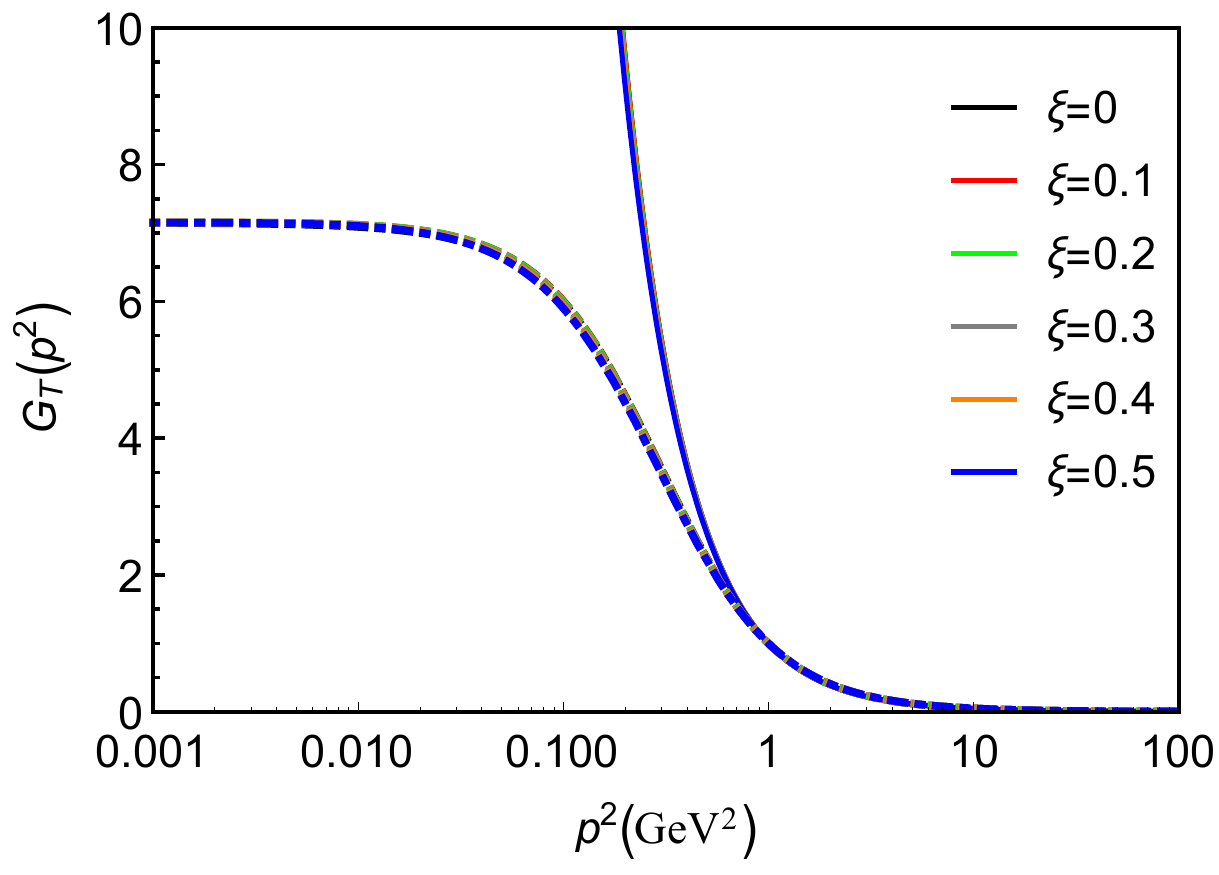}
\caption{ }
 \label{fig:4a}
\end{subfigure}
~
        \begin{subfigure}[h!]{0.312\textwidth}
\includegraphics[width=\textwidth]{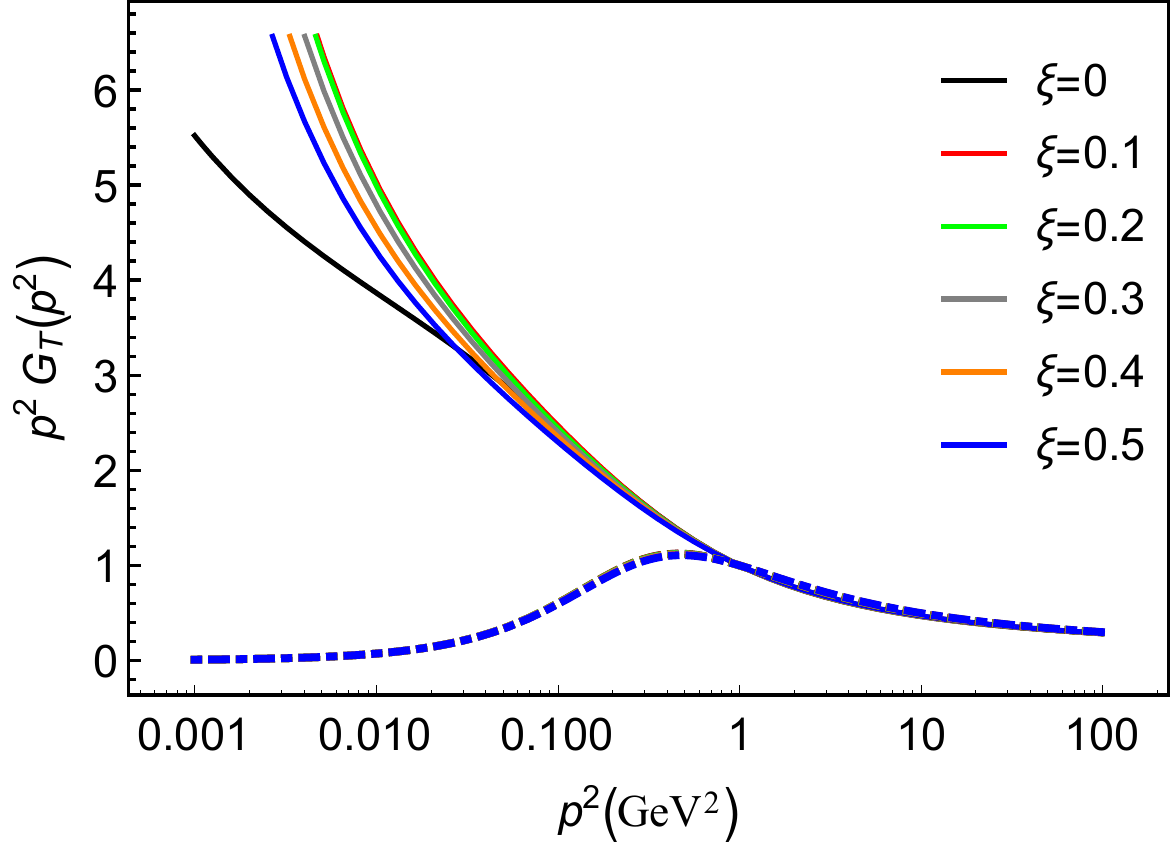}
\caption{ }
 \label{fig:4b}
\end{subfigure}
~
        \begin{subfigure}[h!]{0.321\textwidth}
\includegraphics[width=\textwidth]{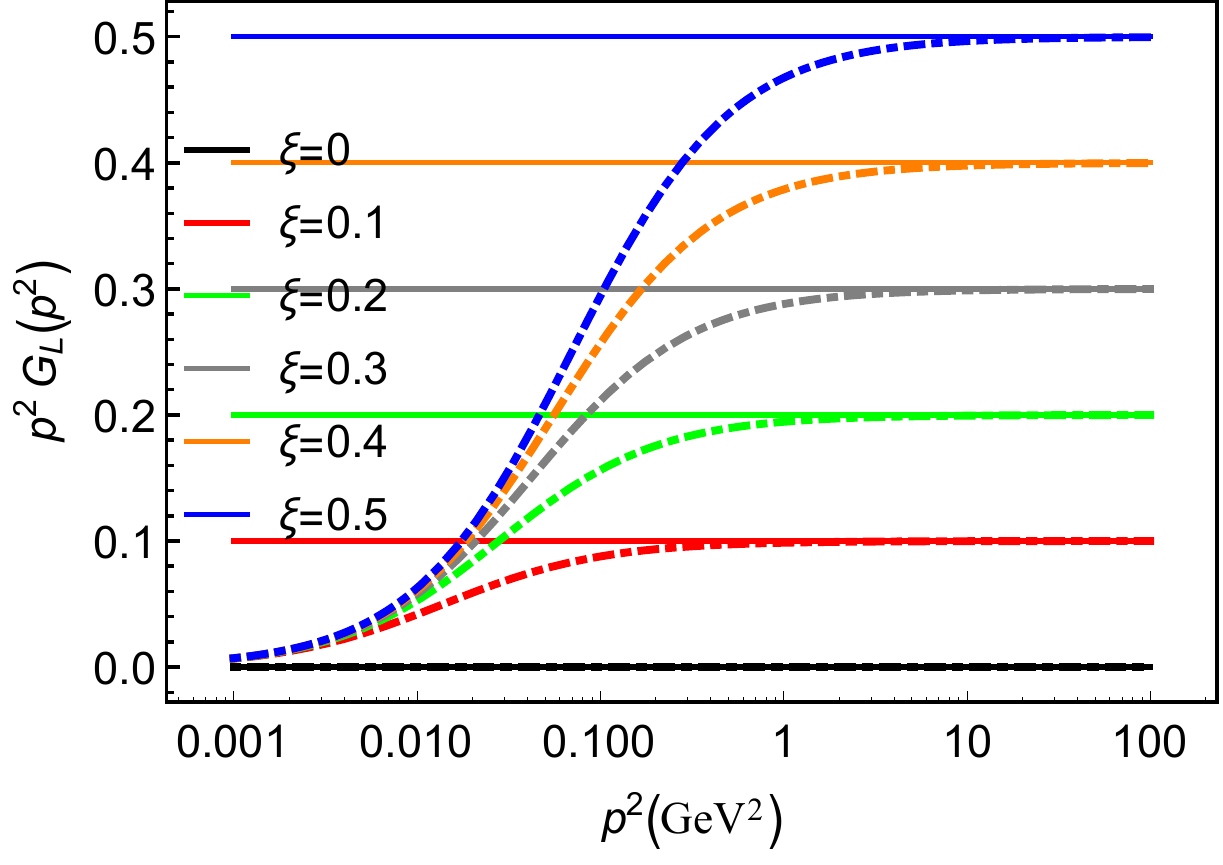}
\caption{ }
 \label{fig:4c}
\end{subfigure}
        \caption{$r=\xi,~\delta_T=0$ BFG transverse gluon propagator (a), transverse dressing function (b), and longitudinal dressing function (c).
Continuous lines correspond to $\lambda=1$, and dashed lines to $\lambda=0$}\label{fig:fig4}
\end{figure*}

In fact, one remarkable result is that $f_L \propto (r-\xi) $: that is, for $r=\xi$, the loop contribution to the BFG self-energy is purely transverse (see Appendix \ref{sec:appbfg}).
Moreover, there is no need for a mass renormalization in this case, so we present in Fig. \ref{fig:fig4} the BFG, $r=\xi$ result for both $\lambda=1$
(continuous lines) and $\lambda=0$ (dashed lines), with $G_T(1\,\text{GeV})=1$ being the only renormalization condition. 

There we see that the corresponding $G_T(p^2)$ for $\lambda=1$ is, however, definitely not adequate. We discuss this matter in the next section.

\subsection{Transversality} \label{sec:transv}

In Fig. \ref{fig:fig4}, we verify that, while the $\lambda=1,\, r=\xi$ BFG result yields a purely transverse self-energy, it leads to an IR enhanced $G_T$, while the corresponding $\lambda=0$ case displays the proper IR finite behavior. While this may be seen as an inconsistency, we recall that,
by keeping $\lambda$ as a parameter, we could consider a massive gluon model approach as well as a NJL-like one.
So, nothing stops us from being more general and, instead of using $(1-\lambda) {\mathcal L}_{\text{se}}$ with Eq.(\ref{eq:lse}),
add the following expression to $\mathcal{L}_{\text{YM}}$:
\begin{widetext}
\begin{equation} \label{eq:lsetwolambdas}
(1-\lambda_T) {\mathcal L}_{\text{se},T}+(1-\lambda_L) {\mathcal L}_{\text{se},L} =
- \frac{1}{2} A^\mu \left( (1-\lambda_T) m^2 \left( g_{\mu\nu} - \partial_\mu \partial_\nu  \partial^{-2} \right) 
+  (1-\lambda_L) m^2 \frac{r}{\xi} \partial_\mu \partial_\nu  \partial^{-2} \right) A^\nu ~.
\end{equation}
\end{widetext}

So, by setting $\lambda_T=0$ and $\lambda_L=1$, the YM Lagrangian plus (\ref{eq:lsetwolambdas}) is a massive gluon model
with a tree-level \emph{transverse mass} for which one employs the NJL-like ``resummation'' for the longitudinal part. Or, if one starts from the local massive model (with $m^2A_\mu A^\mu$), one should have to subtract the longitudinal mass term as further correction to the self-energy.

In other words, it is not by making $r=0$, but by setting $\lambda_L=1$ while $r=\xi$ that the model with a transverse mass gets to have a purely transverse self-energy in the BFG.
For CG, this approach is definitely not sufficient to describe purely transverse gluon mass generation. We shall return to this point in
Section \ref{sec:concluding}.

\begin{figure*}[h!]
\centering
        \begin{subfigure}[h!]{0.315\textwidth}
\includegraphics[width=\textwidth]{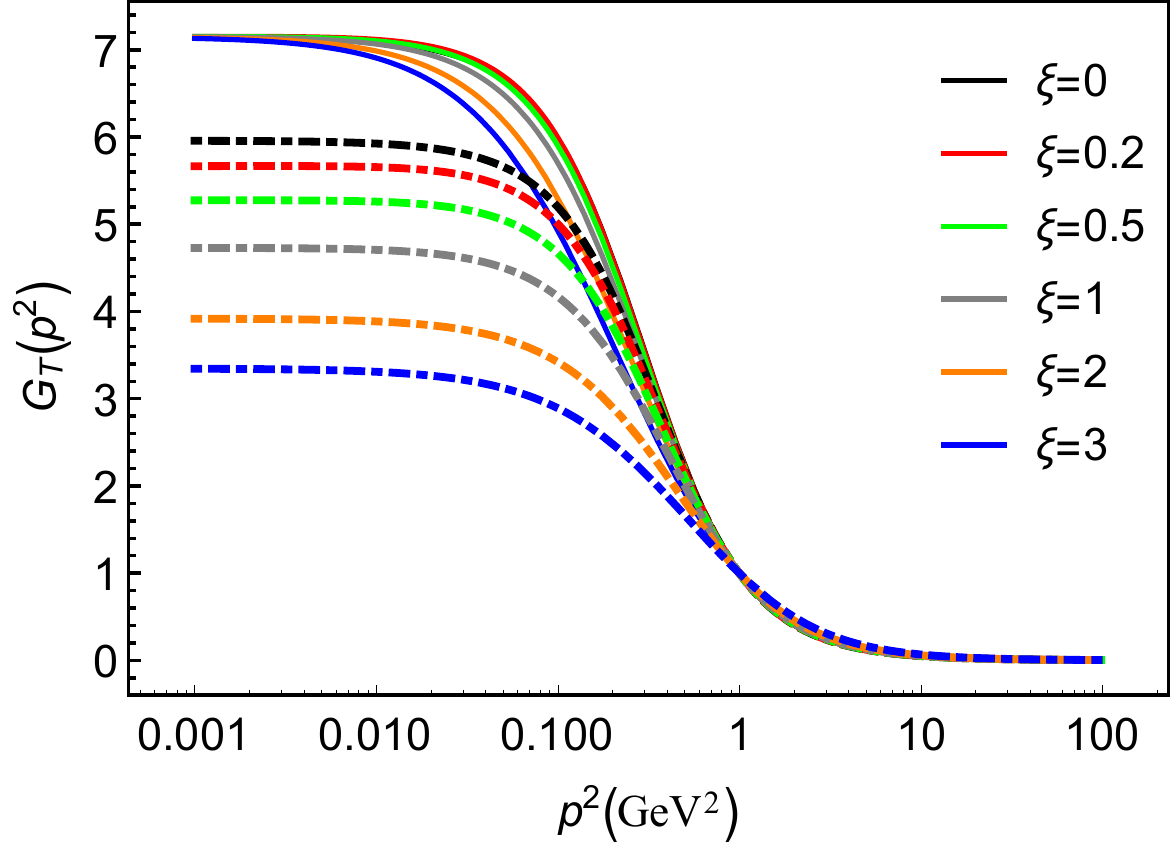}
\caption{}
 \label{fig:5a}
\end{subfigure}
~
        \begin{subfigure}[h!]{0.323\textwidth}
\includegraphics[width=\textwidth]{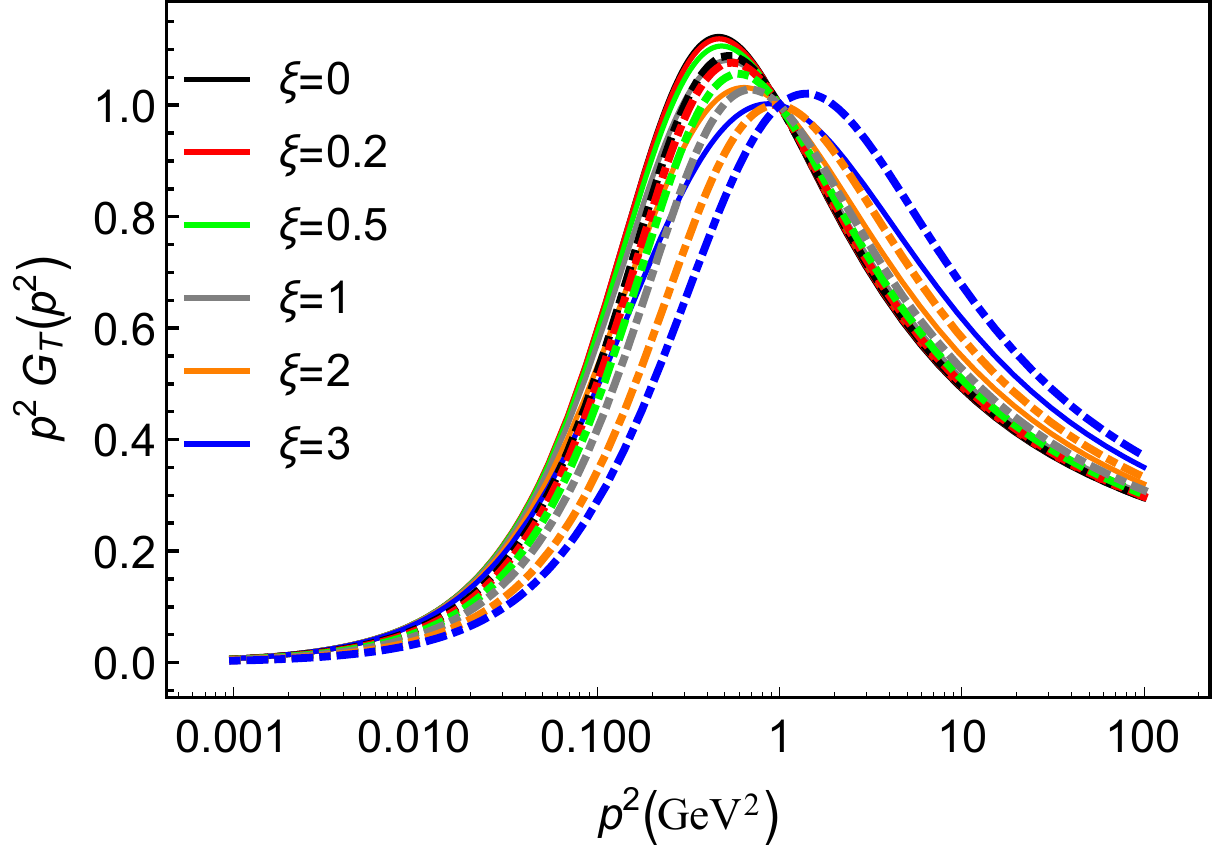}
\caption{}
 \label{fig:5b}
\end{subfigure}
~
        \begin{subfigure}[h!]{0.324\textwidth}
\includegraphics[width=\textwidth]{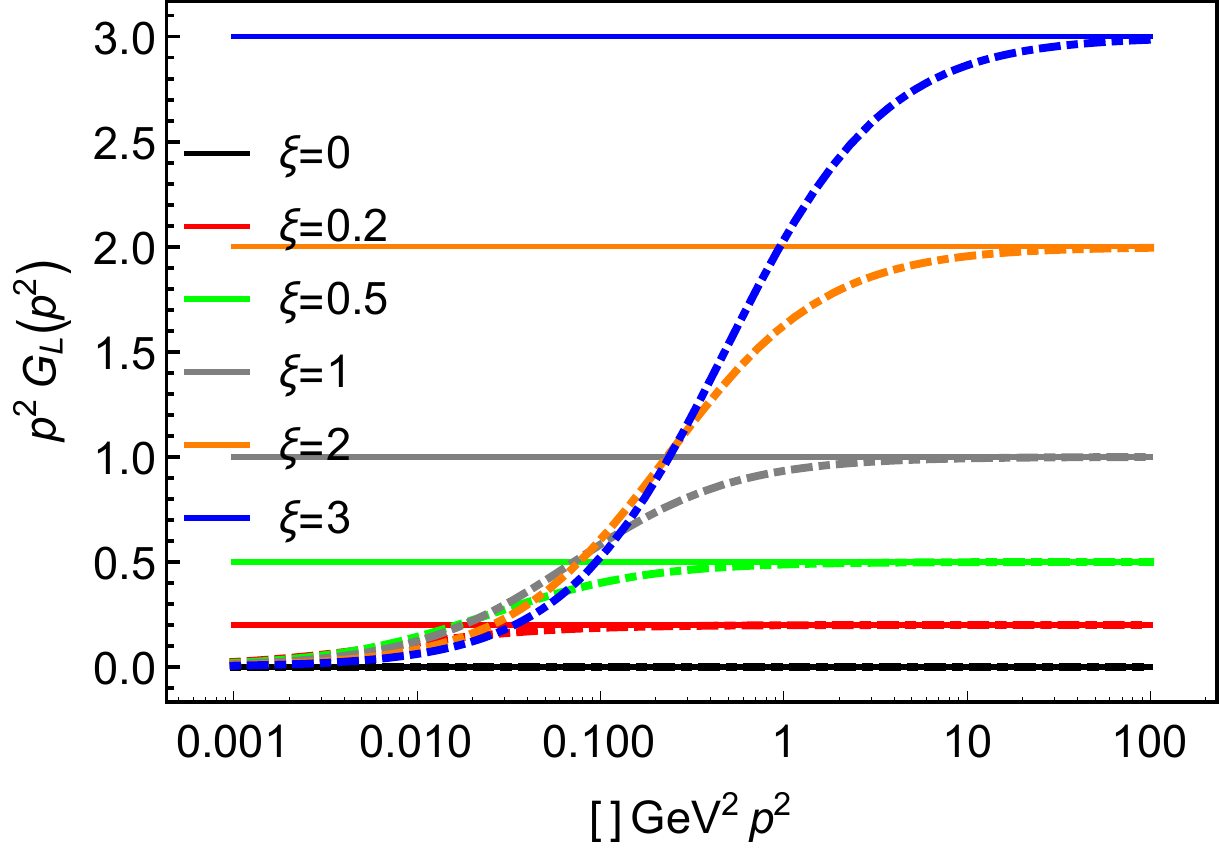}
\caption{}
 \label{fig:5c}
\end{subfigure}
        \caption{BFG results in the $\delta_T=0$ (countinuous lines) and in the (\ref{eq:deltamscheme}) scheme (dashed lines), for the $z_0>0$ parameter set.}\label{fig:fig5}
\end{figure*}

One more aspect we can explore at this point is to compare the $\delta_T=0$ results with the mass-renormalized BFG ones. This is to see how well $G_{T}$ in the scheme of Eq.(\ref{eq:deltamscheme}) 
would approximate the purely transverse $G_{T}$ with $\delta_T=\delta_L=0$. A comparison for the same set of parameters, extrapolated to a wider range in $\xi$,
is shown in Fig. \ref{fig:fig5}.

The point of this comparison is that we still do not have a purely transverse result for the covariant gauges. Suppose one has a more complicated truncation, with dressed vertices, that would yield a purely transverse self-energy. Then one can address the question whether $G_T$ from the former could reasonably approximate the latter. Then a positive answer could be taken to justify neglecting the generated longitudinal self-energy for the purpose of an effective description within certain limits.

We also compare, in Fig.\ref{fig:fig6} below, the two schemes for the $z_0<0$ set of parameters, (\ref{eq:z0neg}), extrapolated for higher values of the gauge parameter. Contrasting Fig. \ref{fig:6a} with Fig. \ref{fig:5a}, we see the large value of $\alpha_s$ leads to a stronger gauge-dependence of the transverse saturation value in the $\delta_T=0$ scheme, while in Fig. \ref{fig:6c} the large coupling effect is evidently problematic, with singularities in the longitudinal dressing function.
The fact that the larger $\alpha_s$ case is problematic, and therefore potentially inaccurate, is related to these results being in plain perturbation theory, and RG improvement might correct these problems and yield more accurate solutions.
The comparison of RG improved quantities shall be presented in a subsequent paper \cite{inprogress}.

\begin{figure*}[h!]
\centering
        \begin{subfigure}[h!]{0.316\textwidth}
\includegraphics[width=\textwidth]{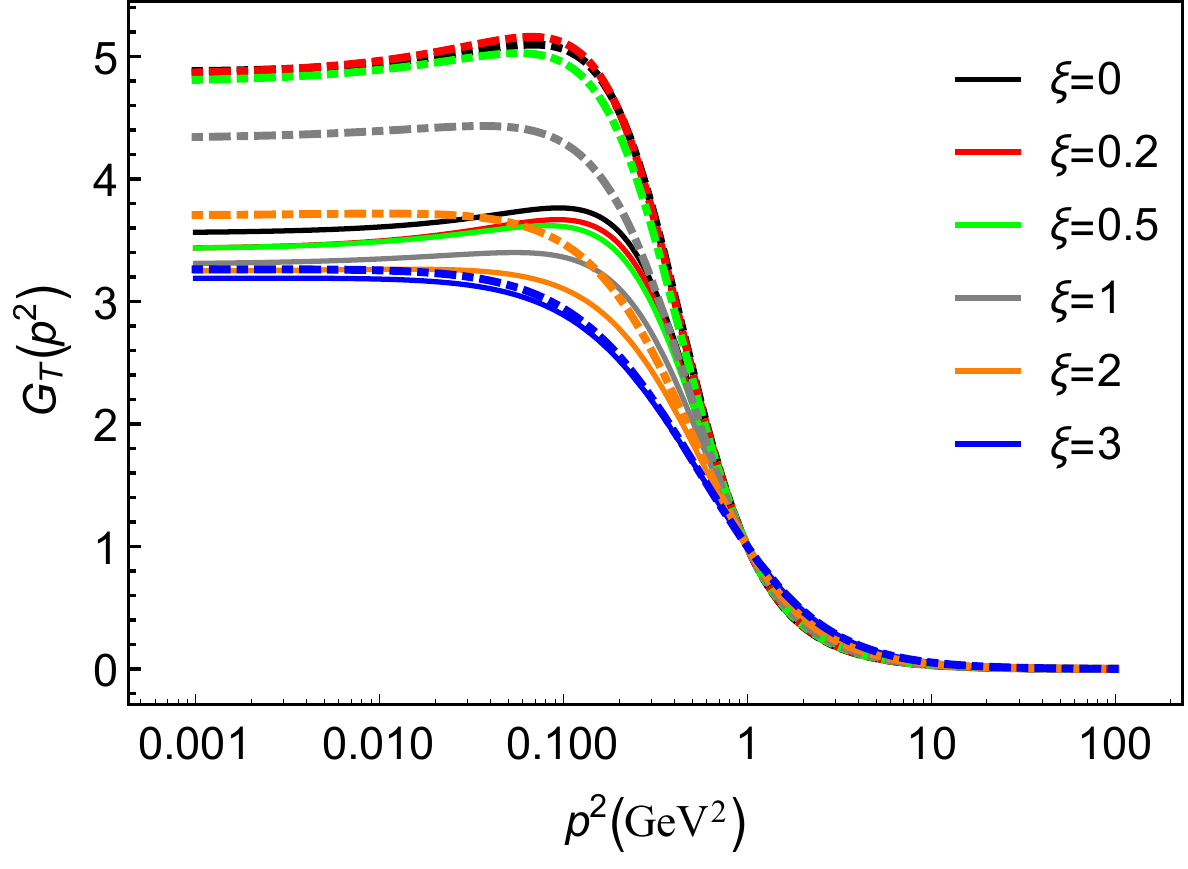}
\caption{}
 \label{fig:6a}
\end{subfigure}
~
        \begin{subfigure}[h!]{0.321\textwidth}
\includegraphics[width=\textwidth]{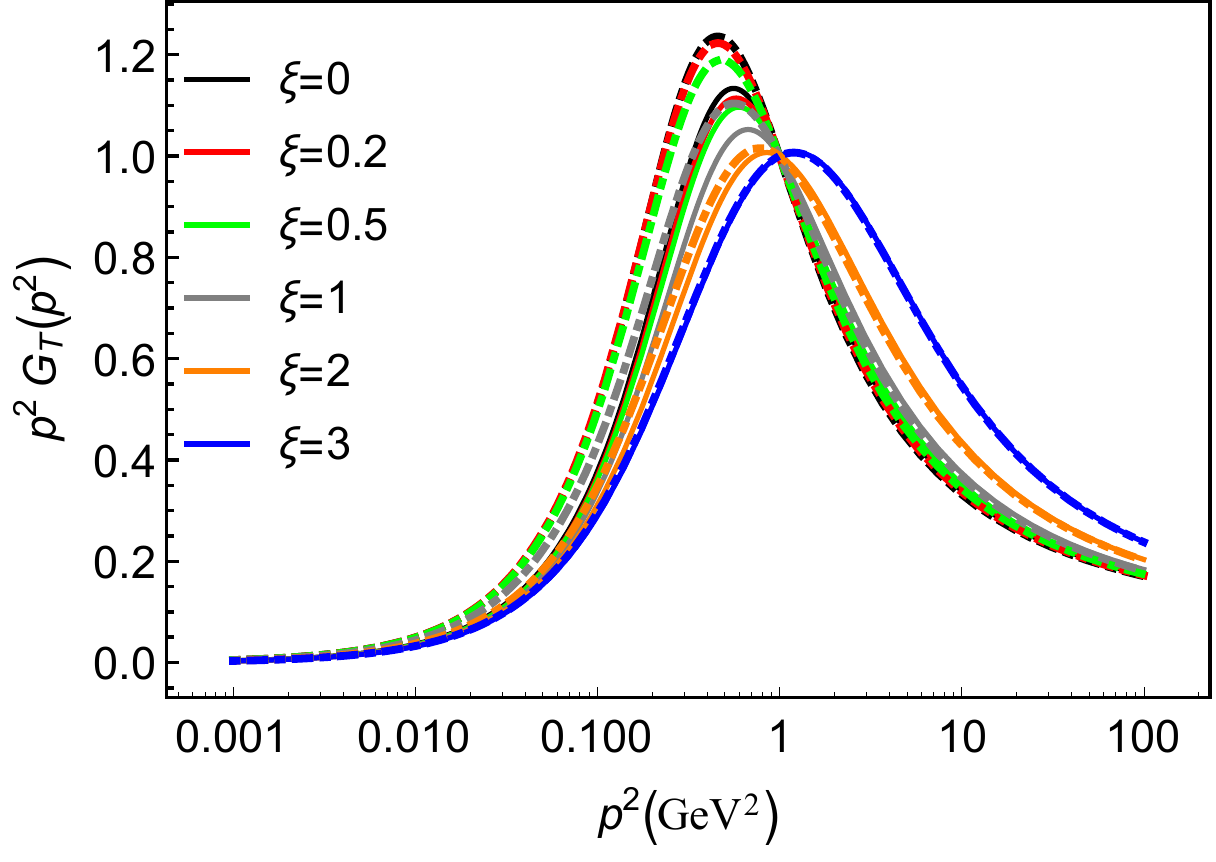}
\caption{}
 \label{fig:6b}
\end{subfigure}
~
        \begin{subfigure}[h!]{0.328\textwidth}
\includegraphics[width=\textwidth]{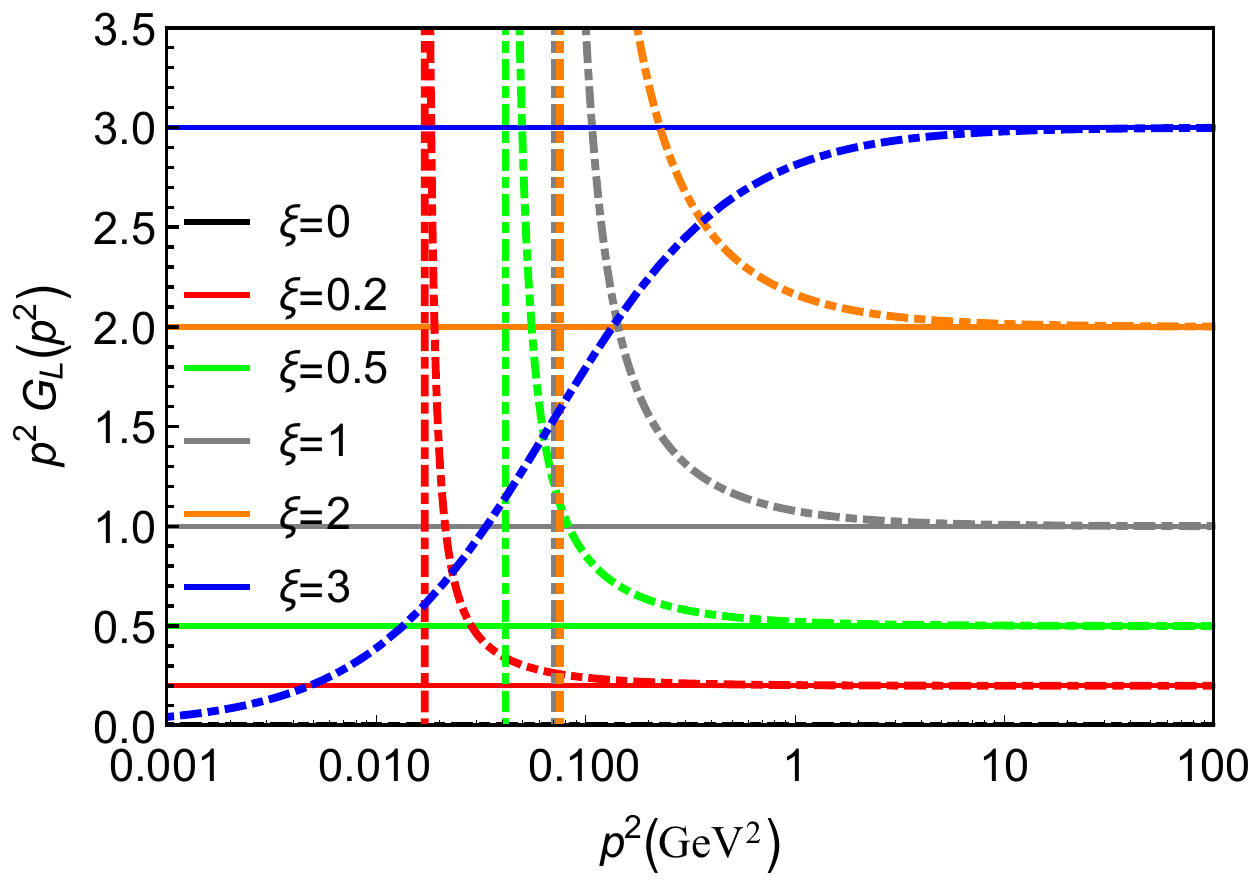}
\caption{}
 \label{fig:6c}
\end{subfigure}
          \caption{BFG results in the $\delta_T=0$ (countinuous lines) and in the (\ref{eq:deltamscheme}) scheme (dashed lines), for the $z_0<0$ parameter set.}\label{fig:fig6}
\end{figure*}

\section{Ghost propagator} \label{sec:ghost}

\begin{figure}[h!]
\centering
        \begin{subfigure}[h!]{0.371\textwidth}
\includegraphics[width=\textwidth]{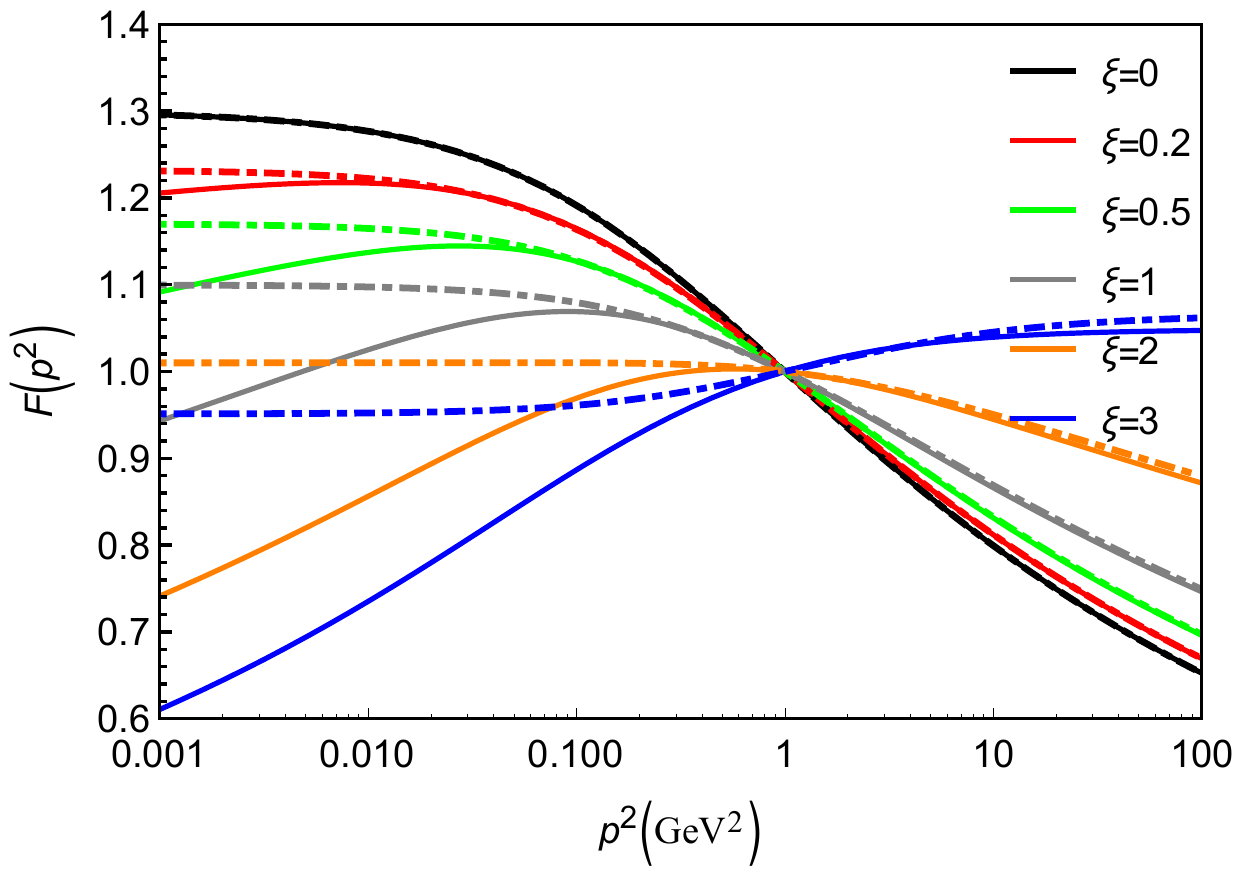}
\caption{Ghost dressing function, with the $z_0>0$ parameter set.}
 \label{fig:7a}
\end{subfigure}
~
        \begin{subfigure}[h!]{0.371\textwidth}
\includegraphics[width=\textwidth]{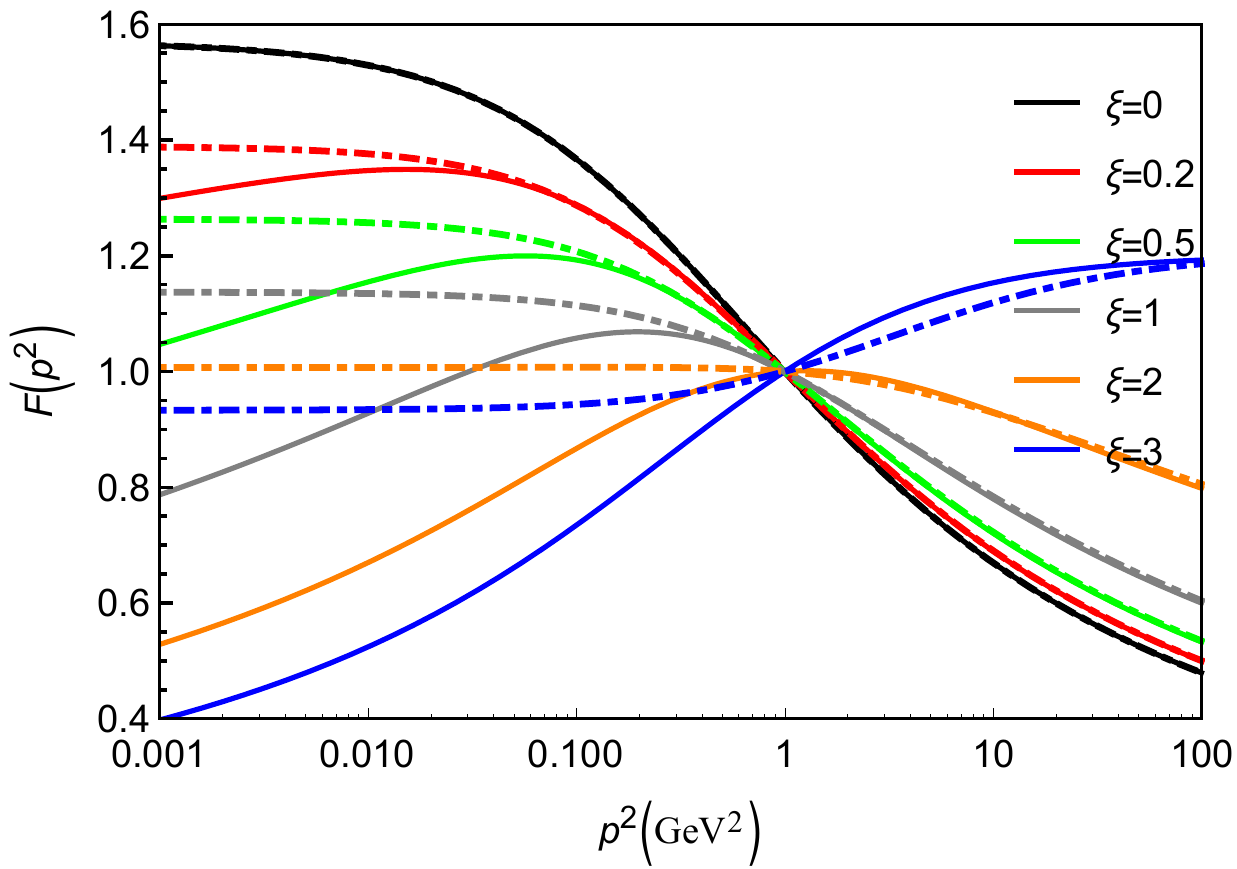}
\caption{Ghost dressing function, with the $z_0<0$ parameter set.}
 \label{fig:7b}
\end{subfigure}
          \caption{Results for the ghost dressing function, for $r=0$ (continuous lines) and $r=\xi$ (dashed lines).}\label{fig:fig7}
\end{figure}

The 1-loop result for the ghost propagator is given by
\begin{equation} \label{eq:gghost}
\tilde{G}^{-1}(p^2) = p^2 \left[ 1 - \frac{N\alpha_s}{16\pi}  f_c(s,r) + \delta_c  \right] =: p^2 F^{-1}(p^2)
\end{equation}
where $F$ is the ghost dressing function, and $f_c$ is given in Appendix \ref{sec:appgh}. Renormalizing in momentum subtraction and setting $F^{-1}(\mu) =: F^{-1}_\mu$ gives
\begin{equation} \label{eq:fghost}
F^{-1}(p^2) = F^{-1}_\mu  - \frac{N\alpha_s}{16\pi} \left[ f_c(s,r) -f_c(\sigma,r) \right] ~,
\end{equation}
where $\sigma:=\mu^2/m^2$. The results for various gauges for $r=0$ and $r=\xi$ are shown in Fig.\ref{fig:fig7}, with $F_\mu =1$ and for both the $z_0>0$ and $z_0<0$ parameter sets.

We note that there is a qualitative distinction between the $r=0$ and $r=\xi$ cases.
While the latter, that corresponds to a local massive Lagrangian, has $\xi>0$ behaving similarly to the Landau gauge, the former, in which the gluon has a nonlocal transverse mass, has $F(p^2)$ decreasing as $p^2$ decreases to the IR.

This difference in asymptotic behavior is to be contrasted with the results from \cite{Aguilar:2015nqa} and \cite{Huber:2015ria} that the deep IR leading behavior is $\xi$-dependent and asymptotically vanishing. 
In fact, Refs. \cite{Aguilar:2015nqa,Huber:2015ria,Huber:2010ne} show that this is connected to the masslessness of the longitudinal component of the gluon propagator, and this connection is manifest in our results.

Concerning the effects of dressing the ghost-gluon vertex, so far what is known is that
its inclusion in the ghost SDE yields
a ghost dressing function that is not IR enhanced in linear covariant gauges \cite{Huber:2010ne}.
The coupled SDEs system for the ghost, gluon, and ghost-gluon vertex in covariant gauges is currently open to investigation,
however the aforementioned connection between the ghost dressing's asymptotic behavior and the masslessness
of the longitudinal gluon propagator supports that this specific feature would not be changed by vertex corrections.

We recall that the BFG transversality result from \ref{sec:bfg} and \ref{sec:transv} involved a transversely massive gluon model with $\lambda_L=1$ but necessarily with $r=\xi$. 
Within the present approach, we can interpret this as consistent with the IR decreasing ghost result in the following way.

Starting from the Lagrangian with $\lambda_T=0$,
\begin{eqnarray}
{\mathcal L} &=& {\mathcal L}_{\text{YM}} + (1-\lambda_T) {\mathcal L}_{\text{se,}T} + (1-\lambda_L) {\mathcal L}_{\text{se,L}} \notag \\
&=& {\mathcal L}_{\text{YM}} + {\mathcal L}_{\text{se,}T} + (1-\lambda_L) {\mathcal L}_{\text{se,L}} ~,  \label{eq:daslagrangian}
\end{eqnarray}
we remark that $-\lambda_L {\mathcal L}_{\text{se,L}}$ as an interaction term is to be present only in the gluon two-point correlator. That is, \emph{the sort of reorganization of the series around a massive gluon self-energy is only meaningful for the gluon propagator itself}.

So, in order to describe correlation functions other than the gluon's two-point while sticking with Lagrangian (\ref{eq:daslagrangian}), one should include the whole $(1-\lambda_L) {\mathcal L}_{\text{se,L}}$ term as part of the leading order gluon propagator.

From (\ref{eq:lsetwolambdas}), we see that this amounts to changing
\begin{equation} \label{eq:ghosteff}
r \mapsto (1-\lambda_L) r
\end{equation}
in our ghost result given in (\ref{eq:ghostr}). Therefore, with $r=\xi$ and $\lambda_L=1$, the result reduces to (\ref{eq:ghost0}), and we obtain the continuous, IR decreasing curves of Fig. \ref{fig:fig7}. 

We remark that the substitution (\ref{eq:ghosteff}) should be done only for correlators other than the gluon's. To be clear, while $r$ is the parameter that determines what form of dressed gluon propagator is to be employed, $\lambda_L$ is the parameter that tells whether or not the longitudinal part of $\mathcal{L}_{\text{se}}$ must be subtracted,
in the NJL-like expansion. Of course one can set $\lambda_L=1$ from the start for those correlators, still $(1-\lambda_L) r m^2$ may serve as an IR regulator when convenient. 

\section{Preliminary Renormalization Group Analysis} \label{sec:rg}

In this section, we briefly sketch the ongoing RG analysis within the present framework and assumptions. Details and further results are left to a later paper \cite{inprogress}, but we show here the straightforward consequences of the self-consistency renormalization scheme and of the transversality result.

\subsection{Self-consistency scheme}

From the nontrivial solution of Eq.(\ref{eq:deltamscheme}), one has that
\begin{equation}
 \delta_m + \delta_Z + \bar{\alpha}  \lim_{s \to 0}s \, f_T(s,r)  = z_\xi + \lambda ~,
\end{equation}
where we wrote $z_0 \mapsto z_\xi$. Then it follows that
\begin{equation} \label{eq:ddmus}
\frac{\mathrm{d}}{\mathrm{d}\mu}Z_m + \frac{\mathrm{d}}{\mathrm{d}\mu}Z_A + \frac{\mathrm{d}}{\mathrm{d}\mu}\left[ \frac{N \alpha_s}{48\pi}\lim_{s \to 0}s \, f_T(s,r) \right] = \frac{\mathrm{d}}{\mathrm{d}\mu}z_\xi ~.
\end{equation}

By construction, $z_\xi$ is given by
\begin{equation}
1+z_\xi = \frac{m_g^2(p^2=0)}{m^2} ~,\notag
\end{equation}
which implies
\begin{equation}
\frac{\mathrm{d}}{\mathrm{d}\mu}z_\xi =  \frac{1}{m^2} \left[    \frac{\mathrm{d}}{\mathrm{d}\mu}m_g^2(0)  - \frac{m_g^2(0)}{m^2} \frac{\mathrm{d}}{\mathrm{d}\mu}m^2 \right] ~.
\label{eq:dzxidmu}
\end{equation}

Therefore, putting (\ref{eq:ddmus}) and (\ref{eq:dzxidmu}) together, we have:
\begin{equation} \label{eq:rgidentity}
F_0 \beta_\alpha + \left( m_g^2(0) - Z_m m^2 \right)\gamma_m + m^2 Z_A \gamma_A
= \frac{\mathrm{d}}{\mathrm{d}\mu}m_g^2(0) - \alpha_s \frac{\mathrm{d}}{\mathrm{d}\mu}F_0~,
\end{equation}
where
\begin{eqnarray}
\gamma_m &=& \frac{\mu}{m^2}\frac{\mathrm{d}m^2}{\mathrm{d}\mu}= - \frac{\mu}{Z_m}\frac{\mathrm{d}Z_m}{\mathrm{d}\mu} ~, \notag \\
\gamma_A &=& \frac{\mu}{Z_A}\frac{\mathrm{d}Z_A}{\mathrm{d}\mu} ~, \notag \\
\beta_\alpha &=& \mu\frac{\mathrm{d}\alpha_s}{\mathrm{d}\mu} ~, \notag \\
F_0 &=& \frac{N}{48\pi}\lim_{p^2 \to 0}p^2 f_T(p^2/m^2,r) ~.\notag
\end{eqnarray}

Expression (\ref{eq:rgidentity}), together with the current knowledge of $\mathrm{d}m_g^2(0)/\mathrm{d}\mu$ from Schwinger-Dyson equations, will be considered in a future paper \cite{inprogress}.

Eventually, having a mass counterterm allows for a perturbative RG treatment and, in the present scheme, leads to a beta function that is explicitly affected by a running mass parameter and by the saturation value of the dynamical gluon mass.

\subsection{Purely transverse Background Field Gauge}

For the purely transverse BFG result, given in Section \ref{sec:transv} and Figs.\ref{fig:fig5} and \ref{fig:fig6}, there is no mass UV divergence, thus allowing the non-renormalization $\delta_m + \delta_Z =: \delta_T \equiv 0$.

Now, from the Slavnov-Taylor identities for the Curci-Ferrari model derived in \cite{Wschebor:2007vh},
one has the following relation:
\begin{equation}
Z_\xi^2 = Z_g \sqrt{Z_A} Z_c = Z_m Z_A Z_c ~,
\end{equation}
where $Z_A$ and $Z_c$ are respectively the gluon and ghost wavefunction renormalization factors, $Z_\xi^2$ the gauge parameter renormalization in the CF model, and 
$$
Z_g :=  \frac{g_{\text{bare}}}{g} ~~,  ~~~ Z_m := \frac{m_{\text{bare}}^2}{m^2} ~.
$$
This implies
\begin{eqnarray}
Z_g &=& 1 + \delta_m +\frac{1}{2}\delta_Z + \mathcal{O}(\alpha_s^2) \notag \\
&=& 1 + \delta_T - \frac{1}{2}\delta_Z + \mathcal{O}(\alpha_s^2)  ~. \label{eq:cfZg}
\end{eqnarray}
So this CF relation, together with $\delta_T$ being identically zero, implies the usual BFG relation \cite{Abbott:1981ke,Binosi:2009qm}
\begin{equation}
Z_g = Z_A^{-1/2} ~,
\end{equation}
meaning all the leading RG logarithms are contained in the gluon wavefunction renormalization, and, as happens in QED, the product $\alpha_s(\mu) G_T(p,\mu)$ defines a RG invariant effective charge \cite{Binosi:2009qm,Aguilar:2010gm}.
Further corrections to it can be considered \cite{Aguilar:2009vn,Aguilar:2010gm}, and again a more complete analysis will follow in future paper \cite{inprogress}.

\section{Conclusions} \label{sec:concluding}

In this paper an effective massive gluon expansion is explored in the context of gluon mass generation in covariant and background field gauges. This expansion has a mass parameter which is renormalized with reference to the saturation of the dynamical gluon mass that can be described in more sophisticated approaches, such as lattice and Schwinger-Dyson equations. Besides finding good agreement with results from those works \cite{Bicudo:2015rma,Aguilar:2015nqa,Huber:2015ria}, we also argue that maintaining transversality of the gluon self-energy is a problem even in Landau gauge, and explicitly show that in covariant gauges neither a diagonal nor a transverse gluon mass can solve this problem within the massive gluon expansion.

We have also shown that it is possible to have a dressed gluon expansion which, at least at next to leading order, yields an identically vanishing longitudinal self-energy for gluons in background field gauges. It should be noted that the BFG presents a correspondence to the Pinch Technique \cite{Aguilar:2006gr,Denner:1994nn}, which so far is the only SDE framework that was shown to lead to a vanishing longitudinal gluon self-energy. Further features of the expansion in the BFG are explored in ongoing work \cite{inprogress}.

The transverse BFG result and the one for the ghost dressing that is
consistent with SDE analyses can be taken as supporting
a specific form of dressed gluon models or expansions for covariant gauges:
one with a transverse gluon mass from the start may be a good IR effective description.
This ultimately corresponds to a nonlocal effective Lagrangian that can be regarded as an approximation of a Lagrangian description of a dynamical gluon mass.

Moreover, the shifting provided by $(1-\lambda_L) {\mathcal L}_{\text{se,L}}$,
whatever the gluon effective ${\mathcal L}_{\text{se,L}}$ would be,
is a NJL-like expansion for the gluon only.
The BFG result tells us then the form of ${\mathcal L}_{\text{se,L}}$,
namely the one in (\ref{eq:lsetwolambdas}), that yields the overall $(r-\xi)$ factor in the longitudinal self-energy, and therefore the desired vanishing correction for $r=\xi$.

It is still unclear whether the longitudinal self-energy that generally arises could be simply factored out in further, phenomenological applications of such massive gluon models. From the knowledge given by SDE studies \cite{Aguilar:2015bud}, it is very likely that dressing vertices is a necessary condition for an effective description to preserve this specific BRST symmetry, however the inclusion of ansatze for vertices within an effective expansion might make it unsuitable for further applications as an improved perturbation theory.  

Nevertheless, this sort of expansion can be employed to probe the infrared behavior of further correlation functions, thus providing some approximate knowledge that may be useful as hints or ansatze in Schwinger-Dyson studies, for instance. 

Finally, the RG analysis, by taking boundary conditions in the range (\ref{eq:z0pos}) or (\ref{eq:z0neg}), may discriminate between the $z_0>0$ and the $z_0<0$ solutions, respectively. This may as well favor one solution or another, or correlate them with certain behaviors for the RG functions within the model.

Work is in progress for obtaining information on vertices in general $R_\xi$ gauges \cite{inprogress2}, as well as for analyzing RG properties in both covariant and background field gauges \cite{inprogress}.


\begin{acknowledgements}
The author would like to thank: O. Oliveira, D. Binosi, P. Bicudo, N. Cardoso, and P.J. Silva for kindly providing their lattice results; E. Swanson for useful discussions and for a critical reading of the manuscript; A. Natale and A. C. Aguilar for discussions and general support; G. Krein, L. Palhares, E. Fraga, M. Guimar\~aes, and S. Sorella for comments on previous related works; and CNPq and CAPES for the financial support through grant number 208188/2014-2.
\end{acknowledgements}

\appendix 

\section{Complete expression of 1-loop results} \label{sec:appendix}

We show our results below, with $2\epsilon = 4-D >0$.

\widetext 
\subsection{Covariant gauges} \label{sec:appcg}

In (\ref{eq:gl}) and (\ref{eq:gt}) respectively, for the case of conventional covariant gauge, we have obtained the following results.

\begin{eqnarray}
f_L(s,r) &=&  \frac{3 (\xi +1)}{s} \left[\frac{1}{\epsilon }+ \log \left(\frac{\mu^2}{m^2}\right) \right]  
+\frac{\xi  (r-2)+5 (\xi +1) s+1}{s^2} \notag \\
&& \notag \\
&& +\log (s) -\frac{(s+1)^3 (r-\xi )}{r s^3} \log (s+1)
+\frac{\xi  (r-s-1) \left(r^2+r (4 s-2)+(s+1)^2\right)}{2 r   s^3} \log (r)
  \notag \\
  && \notag \\
&& -\frac{\xi  \left(r^2+2 r (s-1)+(s+1)^2\right)^{3/2} }{r   s^3}\log \left(\frac{2 \sqrt{r}}{-\sqrt{2 (r+1) s+(r-1)^2+s^2}+r+s+1}\right) \notag \\
&& \notag \\
&& \notag \\
&=& \frac{3 (\xi +1) }{s} \left[ \frac{1}{\epsilon} + \log \left(\frac{\mu^2}{m^2}\right) \right] +\frac{-3 \xi +4 \xi  s+5 s+1}{s^2} \notag \\
&& \notag \\
&& +\log (s) -\frac{(s+1) \left(3 \xi  (s-1)+(s+1)^2\right)}{s^3}  \log (s+1) ~+ \mathcal{O}(r) \notag
\end{eqnarray}

\begin{eqnarray}
f_T(s,r) &=&  \left(-26 + 6 \xi +\frac{9 (\xi +1)}{s} \right) \left[\frac{1}{\epsilon}+ \log \left(\frac{\mu^2}{m^2}\right) \right] -\frac{121}{3}-\xi ^2+\frac{\xi  (9-2 r)+63}{s} +\frac{\xi  (2-r)-1}{s^2}\notag \\
&& \notag \\
   &&  +\left(1-\frac{s^2 (r-\xi )^2}{2 r^2}\right) \log (s) +\frac{\left(s^2-10 s+1\right) (s+1)^3 (r-\xi ) }{r s^3} \log (s+1) + \frac{\xi (r-\xi ) (r+s)^3}{r^2 s}  \log \left(\frac{r}{r+s}\right)  \notag \\
   && \notag \\
&& \!\!\!\!\! -\frac{\xi  \left(r^3 (s+1)^2 (r+3 s-3)+3 r^2 \left(s^4+2 s^3-2  s+1\right)+r \left(s^5+7 s^4+26 s^3+26 s^2+7 s-1\right)-\xi  s^5\right)}{2 r^2 s^3}  \log (r) \notag \\
&& \notag \\
&& \!\!\!\!\!\!\!\!\!\!\!\!\!\!\!\!\!\!\!\!\!\!\!\!\! +\frac{\xi  (s+1)^2 \left(2 (r-5) s+(r-1)^2+s^2\right) \sqrt{r^2+2 r (s-1)+(s+1)^2}}{r s^3} \log \left(\frac{2 \sqrt{r}}{1+s+r-\sqrt{r^2+2 r (s-1)+(s+1)^2}}\right)
  \notag \\
  && \notag \\
&&  +\frac{\xi ^2 \sqrt{s} (s+4 r)^{3/2}}{2 r^2} \log  \left(\frac{2 r}{2r+s+\sqrt{s (s+4 r)}}\right)  +\frac{(s+4)^{3/2} \left(s^2-20 s+12\right)}{2 s^{3/2}}  \log \left(\frac{2}{2+s+\sqrt{s (s+4)}}\right) \notag \\
&&   \notag \\
&& \notag \\
&=& \left(-26 + 6 \xi +\frac{9 (\xi +1)}{s} \right) \left[\frac{1}{\epsilon}+ \log \left(\frac{\mu^2}{m^2}\right) \right] -\frac{121+9 \xi (\xi +3)}{3}+\frac{63}{s}+\frac{3 \xi -1}{s^2}
\notag \\
&& \notag \\
&&+\left(1-\frac{1}{2} s (6 \xi +s)\right) \log (s) +\frac{\left(3 \xi  (s+1) (s-1)^3+(s+1)^3 (s^2-10s+1)\right) }{s^3}\log (s+1)\notag \\
&& \notag \\
&& +\frac{ (s+4)^{3/2} \left(s^2-20 s+12\right)}{2 s^{3/2}} \log \left(\frac{2}{2+s+\sqrt{s (s+4)}}\right) ~+ \mathcal{O}(r) \notag
\end{eqnarray}

\subsection{Background field gauges} \label{sec:appbfg}

In (\ref{eq:gl}) and (\ref{eq:gt}) respectively, for the case of background field gauge, the results are the following.

\begin{eqnarray}
f_L(s,r) &=&  (r-\xi ) \left\{ -\frac{3 (\xi -1) }{\xi  s} \left[ \frac{1}{\epsilon} + \log \left(\frac{\mu^2}{m^2}\right) \right]   -\frac{(-2 \xi +(\xi +2) r+5 (\xi -1) s-1)}{\xi  s^2}  \right.  \notag \\
&& \notag \\
&&  -\frac{(s+1)^3}{r s^3} \log (s+1)
   -\frac{(r+s)^3}{\xi  s^3} \log \left(\frac{r}{r+s}\right) \notag \\
   && \notag \\
&& +\frac{\left(r \left(3 \left(r^2-1\right) s+3 (r-1) s^2+(r-1)^3+s^3\right)-\xi  (r-s-1) \left(r^2+r (4 s-2)+(s+1)^2\right)\right)}{2 \xi  r s^3} \log (r) \notag \\
&& \notag \\
&& \left. -\frac{(r-\xi ) \left(r^2+2 r (s-1)+(s+1)^2\right)^{3/2} }{\xi  r s^3} \log \left(\frac{2 \sqrt{r}}{-\sqrt{2 (r+1) s+(r-1)^2+s^2}+r+s+1}\right) \right\} \notag \\
&& \notag \\
&& \notag \\
&=& \frac{3 (\xi -1) }{s} \left[ \frac{1}{\epsilon} + \log \left(\frac{\mu^2}{m^2}\right) \right]
-\frac{3 \xi -4 \xi  s+5 s+1}{s^2} -\log (s)
+\frac{\left((s+1)^3-3 \xi \left(s^2-1\right)\right)}{s^3} \log (s+1) \notag \\
&& \notag \\
&&  ~+ \mathcal{O}(r) \notag
\end{eqnarray}

\begin{eqnarray}
f_T(s,r) &=& \left( -44 -\frac{9 (r-\xi ) (\xi -1)}{\xi  s} \right) \left[ \frac{1}{\epsilon} + \log \left(\frac{\mu^2}{m^2}\right) \right] -\frac{223}{3}  -\xi  (\xi +14)+\frac{9 \xi +\left(\frac{17}{\xi }+7\right) r+31}{s} \notag \\
&& \notag \\
&&+\frac{(r-\xi ) (-2 \xi +(\xi +2) r-1)}{\xi  s^2}  + \left(4-\frac{s^2 (r-\xi )^2}{2 r^2}\right) \log (s)  \notag \\
&& \notag \\
&&+\frac{(r-\xi )(s+1)^3 \left(s^2-10 s+1\right) }{r s^3} \log (s+1) +\frac{(r-\xi ) (r+s)^3 (r-\xi  s)^2 }{\xi  r^2 s^3}\log
   \left(\frac{r}{r+s}\right) \notag \\
   && \notag \\
&&  \!\!\!\!\!\!  -\frac{1}{2 \xi  r^2 s^3} \left[-\xi ^3 s^5+ r^6-r^5 (2 \xi +(2 \xi -3) s+3)+r^4 \left(\xi ^2+6 \xi +\left(\xi ^2-6 \xi +3\right) s^2+2 \left(\xi ^2-6\right) s+3\right)  \right. \notag \\
&& \notag \\
&& \!\!\!\!\!\!  +r^3 \left(-3 \xi ^2-6 \xi
   +\left(3 \xi ^2-6 \xi +1\right) s^3+3 \left(\xi ^2+3\right) s^2-3 \left(\xi ^2-6 \xi -3\right) s-1\right) \notag \\
   && \notag \\
&& \!\!\!\!\!\!\!\!\!\!\!\!\!\!\!\!\!\!\! \left.+\xi  r^2 \left(3 \xi +(3 \xi -2) s^4-2 (3
   \xi +7) s^3-36 s^2-2 (3 \xi +8) s+2\right)+\xi ^2 r \left(s^5+7 s^4+26 s^3+26 s^2+7 s-1\right)  \right] \log (r) \notag \\
&& \notag \\
&&
\!\!\!\!\!\!\!\!\!\!\!\!\!\!\!\!\!\!\!\!\!\!\!\!\!\!\!\!\!\!\!\!\!\!\!\!\!\!\!\!\!\!\!\!\!\!\!\!\!\! +\frac{\left(r^2+2 r (s-1)+s^2-10 s+1\right)
   \sqrt{r^2+2 r (s-1)+(s+1)^2} (r-\xi  (s+1))^2}{\xi  r
   s^3} \log \left(\frac{2 \sqrt{r}}{1+r+s-\sqrt{r^2+2 r (s-1)+(s+1)^2}}\right) \notag \\
&& \notag \\
&& \!\!\!\!\!\!\!\!\!\!\!\!\!\!\!\!\!\!\!\!\!\!\!\!\!\!\!\!\!\!\!\!\!\!\!\!\!\!\!\!\!\!\!\!\!\!\!\!\! + \frac{(4 r+s)^{3/2} (2 r-\xi  s)^2}{2 r^2 s^{3/2}}  \log \left(\frac{2 r}{2r+s+\sqrt{s (4 r+s)}}\right)  + \frac{(s+4)^{3/2} \left(s^2-20 s+12\right)}{2 s^{3/2}} \log \left(\frac{2}{2+s+\sqrt{s (s+4)}}\right) \notag \\
&& \notag \\
&& \notag \\
&=& \left( -44 -\frac{9 (1-\xi)}{s} \right) \left[ \frac{1}{\epsilon} + \log \left(\frac{\mu^2}{m^2}\right) \right]-\frac{223}{3} -3 \xi  (\xi +7) +\frac{31}{s} +\frac{3 \xi +1}{s^2} \notag \\
&& \notag \\
&& -\frac{s^3+(6 \xi -4) s^2-6 (2 \xi +1) s}{2s} \log (s) +\frac{\left(s^2-1\right) \left(3 \xi +s^3+3 (\xi -3) s^2-3 (2 \xi +3) s+1\right)}{s^3}  \log (s+1)\notag \\
&& \notag \\
&& +\frac{(s+4)^{3/2} \left(s^2-20 s+12\right)}{2 s^{3/2}} \log \left(\frac{2}{2+s+\sqrt{s (s+4)}}\right)  ~+ \mathcal{O}(r) \notag
\end{eqnarray}

\subsection{Ghost propagator} \label{sec:appgh}

Finally, in (\ref{eq:gghost}) and (\ref{eq:fghost}) for the ghost correlator, the result is:

\begin{eqnarray}
f_c(s,r,m_T) &=& (3-\xi ) \left[\frac{1}{\epsilon }+\log \left(\frac{\mu^2}{m^2}\right)\right] +5 -\xi +\frac{1+\xi  r}{s}   \notag  \\
&& \notag \\
&&  +\frac{\xi  r (r+s)}{s^2} \log (r)
+\frac{\xi  (s-r) (r+s)^2 }{r s^2}\log \left(r+s\right)
+\frac{s (r-\xi) }{r}\log \left(s\right)
-\frac{(s+1)^3 }{s^2} \log \left(s+1\right) \label{eq:ghostr} \\
&& \notag \\
&=&  (3-\xi ) \left[\frac{1}{\epsilon }+\log \left(\frac{\mu^2}{m^2}\right)\right] +5 +\frac{1}{s} +(\xi +s) \log \left(s\right) -\frac{(s+1)^3}{s^2}  \log \left(s+1\right) ~+ \mathcal{O}(r) \label{eq:ghost0} ~.
\end{eqnarray}

\bibliographystyle{apsrev4-1}   
\bibliography{Bib1}  

\end{document}